\documentclass{article}
\usepackage{geometry}
\usepackage[utf8]{inputenc}
\usepackage{graphicx}
\usepackage{amsmath}
\usepackage{amsfonts}
\usepackage{caption}
\usepackage{natbib}
\usepackage{xcolor,colortbl}
\usepackage{setspace}
\usepackage{authblk}
\newtheorem{definition}{Definition}

\setcitestyle{authoryear}

\title{Quantifying the Coherence of Development Policy Priorities}

\author[1,2]{Omar A. Guerrero}
\author[3]{Gonzalo Casta\~neda}
\affil[1]{The Alan Turing Institute, London}
\affil[2]{Department of Economics, University College London, London}
\affil[3]{Centro de Investigación y Docencia Econ\'omica (CIDE), Mexico City}
\date{}

\begin{document}

\maketitle

\begin{abstract}
    Over the last 30 years, the concept of policy coherence for development has received especial attention among academics, practitioners and international organizations. However, its quantification and measurement remain elusive. To address this challenge, we develop a theoretical and empirical framework to measure the coherence of policy priorities for development. Our procedure takes into account the country-specific constraints that governments face when trying to reach specific development goals. Hence, we put forward a new definition of policy coherence where context-specific efficient resource allocations are employed as the baseline to construct an index. To demonstrate the usefulness and validity of our index, we analyze the cases of Mexico, Korea and Estonia, three developing countries that, arguably, joined the OECD with the aim of coherently establishing policies that could enable a catch-up process. We find that Korea shows significant signs of policy coherence, Estonia seems to be in the process of achieving it, and Mexico has unequivocally failed. Furthermore, our results highlight the limitations of assessing coherence in terms of naive benchmark comparisons using development-indicator data. Altogether, our framework sheds new light in a promising direction to develop bespoke analytic tools to meet the 2030 agenda.
    
    

\end{abstract}

\newpage

\section{Introduction}

When governments aim at improving specific socioeconomic indicators, they formulate and implement policies that are often paved with inefficiencies of various kinds. In addressing policy design and implementation, academics and development consultants often highlight the importance of coherence, acknowledging the fact that development goals and policies are multidimensional. Nevertheless, the term `policy coherence' is a loosely defined term that has different meanings across various researchers and organizations. The need for conceptual clarity and unambiguous measurements calls for a redefinition of the concept. In this paper, we develop such definition and construct a relevant metric. The proposed index allows estimating how coherent are the policy priorities of a country when it attempts to reach a specific set of development goals. 

Traditionally, assessing policy coherence involves qualitative methods such as analyzing official speeches and documents that signal how aligned are certain transformative policies in relation to a set of established goals \citep{oecd_better_2015,oecd_better_2016,oecd_policy_2017,oecd_policy_2018}. While these approaches may give us initial pointers, there is still a long way to in measuring policy coherence with certain degree of confidence. This is so because policy priorities of governments are not directly observable through discourse or coarse-grained public expenditure data. In other words, qualitatively evaluating coherence can be extremely misleading due to the political economy that shapes factual policy priorities.

The construction of a metric that quantifies policy coherence is paramount for several reasons. First, it provides a less discretionary way to measure how committed a government is to reach certain development targets. Second, it allows comparisons between countries and regions, which is extremely helpful to evaluate and rethink international development agendas. Third, by relying more on data and less on highly-specialized qualitative expertise, a quantitative metric can relax the constraint of scarce technical capabilities to which many developing countries are subject. Fourth, it helps governments designing timely responses by informing them on how to reorganize their policy priorities. Fifth, it reinforces the need for evidence-based policymaking towards the United Nations' 2030 Agenda with respect to the Sustainable Development Goals (SDGs).  

Despite the potential benefits of a coherence metric, there do not exist scalable and robust indices. This is caused by different data- and theoretic-related challenges that need to be overcome. Here, we mention a few. First, in order to remove the veil of non-observable policy priorities, it is necessary to model the policymaking process that gives place to observed development-indicator data. This demands modeling tools where the micro-incentives of the relevant actors generate the macro-behavior of the indicators. Second, in order to identify the long-discussed synergies and trade-offs between public policies, new statistical methods are necessary. In particular, network-estimation methods need to be tailored to the coarse-grained quality of development-indicator data. Third, when modeling the policymaking process, it is crucial to account for the institutional context of the country under study; in particular, for mechanisms related to inefficiencies such as a poor governance and corruption. Fourth, policy coherence is intimately related to the specific context of each country. Hence, empirical analyses that pool cross-national data run the risk of neglecting relevant contextual features. Fifth, it is also necessary to estimate the policy priorities that countries would establish, should they decide to pursue those goals. This is so because these `counterfactual priorities' provide a reference point to construct a normalized measurement that takes into account the specific constraints and inefficiencies that countries face. Overall, the challenges in building a coherence index are numerous and difficult to tackle. Our work provides a first step in trying to overcome some of them.

Implicitly, our definition of policy coherence considers that development indicators are the consequence of government expenditures, implementation inefficiencies and spillover effects. Therefore, the corresponding metric requires inferring factual policy priorities from a political economy game on a network. For this purpose, we use a recently developed framework called \textit{Policy Priority Inference} (PPI) \citep{castaneda_how_2018}, which simulates the policymaking process and allows estimating the government's allocation profile. Using PPI, and with a given country and goals, we measure coherence through the discrepancy between the estimated policy priorities (retrospective analysis) and the priorities that the government would establish if it were serious to pursue those goals.

In order to empirically identify development goals, we frame our application in the context of the Organization for Economic Cooperation and Development (OECD). This is a convenient setting to study policy coherence since, arguably, membership to this intergovernmental entity is conditioned on certain degree of alignment to a set of principles established by the incumbent members. Therefore, we argue that post-1990 member countries had incentives to follow the lead of the early OECD members. The core of our analysis concentrates on Mexico, a developing economy that, through official government discourse, has claimed the need to embrace the implementation of coherent policies.\footnote{\textit{``A National Council for the 2030 Agenda for Sustainable Development, chaired by the president, was established in 2017. Its main purpose is to coordinate the actions for the design, execution and evaluation of [...] policies [...] for the compliance with the 2030 Agenda''} \citep[p.~135]{oecd_policy_2018}. The website www.gob.mx/agenda2030 contains a repository of documents and information regarding SDGs in Mexico.} In fact, this discourse is not exclusive of Mexico's last administration, but it has been a defining feature all Mexican governments in of the last three decades. In spite of this apparent enthusiasm for catching up with the OECD early members, our findings suggest that Mexico's policy priorities have not been coherent. Even worse, we find that Mexico's priorities have been quite the opposite to what they would be if the government was serious about reaching these goals.

The rest of the paper is structured in the following way. In section \ref{sec:coherence}, we discuss the concept of coherence with regard to policy priorities and present a literature review on alternative approaches. Section \ref{sec:methods} presents the data and methods to estimate policy priorities through the PPI framework. In section \ref{sec:results}, we build the coherence index and present our main empirical findings. Finally, in section \ref{sec:conclusions} we discuss the limitations and potentials of our approach, and provide some conclusions.

\section{On the coherence of policy priorities}\label{sec:coherence} 

The idea of policy coherence for development has lingered in academic research and policy reports for a few decades. While development economists have not been particularly interested in this concept (perhaps because their main concern focuses on growth and income distribution), a large number of practitioners and academics in the broader field of development studies have extensively discussed the idea. Broadly speaking, coherence has been qualitatively studied under the literature of Policy Coherence for Development (PCD) \citep{forster_policy_2013}. Originally, PCD was conceived as a principle for the international aid system. The main idea was that donor countries should also consider the impact that policies established for their own benefit have on the development of poor countries, and not only the effects on development aid policies in those nations \citep{sianes_shedding_2017,barry_policy_2010}. PCD, however, has evolved to the point of becoming the evaluation standard for the planning and implementation of policies in any country trying to achieve a set of sustainable development goals, irrespective if it is an advanced, emerging or poor nation. Arguably, one of the main drivers in the widespread use of this standard has been the OECD through its several reports on policy coherence. For instance, the opinion expressed by Angel Gurr\'ia --Secretary-general-- in the foreword of its 2018 report (\citeauthor[p.~3]{oecd_policy_2018}) provides an up-to-date notion of poliocy coherence:

\vspace{5mm}

``\textit{(PCD) calls for breaking out of policy silos and increasing capacities to identify, understand and manage interactions and interconnections among SDGs. It entails harnessing synergies, managing trade-offs and policy conflicts, and addressing the potential transboundary and intergenerational policy effects of domestic and international action}''.

\vspace{5mm}

This definition constitutes an important step in the right direction for the PCD framework because it makes explicit the importance of policy-policy interdependencies and policy-goal interactions. With such high-level official recognition, a wide spectrum of development analysts have re-framed PCD as a systemic problem. In consequence, the to identify potential reinforcing and conflicting effects between development indicators has become prevalent across international organizations and academics. For example, when the OECD and other multilateral agencies (\textit{e.g.}, UNDP and the World Bank) extended their agenda from PCD to policy coherence for sustainable development (PCSD) in 2014, they aimed at integrating economic, social and environmental dimensions of development across all levels of domestic and international policymaking through their complex interdependencies. Unfortunately, this demand for identifying interactions between policies has exposed severe technical limitations in qualitative approaches; in particular, their heavy dependence on expert knowledge in highly-specific fields, which precludes scalability and introduces conflicting biases. Therefore, developing systematic, quantitative and scalable frameworks has become an unavoidable endeavour. 

Before diving into the proposed method, the reader should be aware that the development literature defines policy coherence at different levels and stages (see, for instance, \cite{curran_policy_2018,carbone_mission_2008}). First, horizontal coherence alludes to the interactions between policy issues and how these make possible attaining different goals simultaneously. Vertical coherence is used to describe the connections between policies at different government levels (\textit{e.g.}, regional and national, national and supra-national). Second, policy coherence can be analyzed at two stages: design and implementation. The former relates to the formulation of policy priorities by analysts and policymakers. The latter involves the coordination of different government actors responsible for the operational side of policies. In both classifications, assessing policy coherence requires a partnership between scientists of different kinds and technocrats dealing with public-administration issues. 

In this paper, we are interested in describing an approach that emphasizes policy coherence at the horizontal level. This proposal is intended for the design stage and it is based on modern scientific tools for data analysis. In this sense, our focus is more narrowly defined than PCD, which is an overarching term for different discussions on policy coherence. Hence, in order to prevent any confusion with the OECD definitions, we will avoid using the term PCD altogether. Instead, we use the term \textit{coherence of policy priorities} (or, generically, policy coherence) to refer to the analysis derived from our definition. Delimiting the scope of our study allows us to construct a more comprehensive definition of coherence. One that goes beyond the identification of positive and negative spillovers between policy issues, and helps us to consider the specific constraints and inefficiencies that nations face during their development.

\subsection{Some challenges in measuring policy coherence and related literature}

In this section, we discuss some of the main challenges that need to be met in order to quantify policy coherence. We elaborate on five problems that, in our opinion, are not properly dealt with by existing approaches. In addition, we believe that each of these challenges is addressed by the Policy Priority Inference framework.

The first problem --implementation inefficiencies-- relates to the limitation of directly observing coarse-grained government-expenditure data, but not the policy priorities behind official statistics and discourse. The second --spillovers effects-- refers to the problem with inferring coherence from the interdependencies of socioeconomic-indicators. The third issue --network estimation-- consists of the empirical problem of inferring interactions between development goals (or indicators) via qualitative tools (\textit{e.g.}, expertise and stakeholders information). The fourth problem --context specificity-- relates to the loss of a country's contextual information when pooling cross-national data for statistical analysis. Finally, the fifth challenge --implicit benchmark-- alludes to the need of counterfactual analyses in order to generate country-specific reference points. Next, we proceed to discuss each challenge in detail. At the same time, we review some of the existing methods, highlighting some of their virtues and pitfalls.

\subsubsection{Implementation inefficiencies}

A recent study on public expenditure by the Inter-American Development Bank reveals a major problem of resource misusage in Latin-America. This wastage is the result of lacking of professionalism in the bureaucracy, negligence, corruption or a combination of all these factors \citep{izquierdo_better_2018}. For instance, the study estimates that, on average, inefficiencies in just three policy topics (procurement, civil services and targeted transfers) account for 4.4\% of the regional GDP, and about 16\% of the total government expenditure (p. 63). To put this into perspective, note that a similar amount of expenditure with respect to GDP is, on average, allocated to health (4.1\%) and education (4.8\%). Besides these technical inefficiencies, there are important allocative inefficiencies arising from a poor distribution of resources across generations, government levels and policy issues. 

A comprehensive measure of policy coherence should consider both technical and allocative inefficiencies (being corruption an important component of the former term). In our view, the biggest flaw of the current frameworks is the omission of the policy-making process and, thus, the assumption that expenditure data reveal the true government's intentions on how development targets will be reached. In reality, political economy considerations play an important role on determining how many resources are detoured for personal gains or how they are wasted by bureaucratic inefficiencies. Consequently, governments adapt their budgets to the political economy, obfuscating the connection between their true priorities and public expenditure data. Furthermore, these dynamics preclude the estimation of policy priorities from single-period data.\footnote{This situation worsens if the data cannot be properly disaggregated into fine-grained types of expense (transformative or committed), policy issues (topics) and government sources (federal, state and municipal).}

Clearly, technical and allocative inefficiencies constrain development and, hence, the feasibility to reach a development targets. For this reason, it is necessary to model the adaptive process through which the government evolves its priorities while restricted by these inefficiencies. It is important to emphasize that priorities inferred through such models not only are indicative of factual government intentions, but also of the budgetary adjustments due to corruption dynamics.

An early attempt to get closer to process-models comes from Systems Dynamics \citep{pedercini_dynamic_2010,collste_policy_2017}. These models simulate the transformation of socioeconomic indicators as they evolve through time under alternative budgetary allocations. They show how the initial distribution of resources impact directly on the targeted indicators, and indirectly on the evolution of other indicators through a map of stocks-and-flows linkages. The stocks-and-flows nature of these models make then quite aggregate. Thus, it is not possible to disentangle the policymaking process (and the political economy) that takes place within each stock and flow. This makes them difficult to validate and unable to account for the enormous amount of resources that never make it to their final destination; something critical in developing countries. 

More obvious shortcomings exist in static models that estimate spillover networks through subjective/qualitative procedures \citep{leblanc_integration_2015,weitz_systemic_2018,allen_prioritising_2018}, conventional statistical techniques \citep{pradhan_systematic_2017,ceriani_multidimensional_2016,cinicioglu_exploring_2017,czyzewska_bayesian_2014}, a combination of the previous two \citep{zhou_sustainable_2017} or co-occurrence methods \citep{el-maghrabi_sustainable_2018}. The problem with these studies is that they do not attempt to estimate policy priorities. Instead they assume that synergies and trade-offs between indicators can be directly mapped into policy priorities. Thus, they ignore the discrepancy between the design and the implementation of public policies. Presumably, a rule of thumb for an ex-post evaluation of policy coherence for this type of analyses, would check if the centrality of policy issues (defined as nodes in the network) correlates with the share of government expenditure devoted to that issue. As suggested above, this would be misleading since a large portion of the observed expenditure may be inefficiently used or wasted in corruption.

\subsubsection{Spillover effects}

Nowadays, multilateral organizations acknowledge that development goals are part of an `indivisible whole' and, thus, they advocate for policy coherence when planning is carried out. In order to perform this task, the first generation of systemic studies associates coherence with the promotion of policies whose indicators show synergistic effects (positive spillovers). It also discourages investing on issues that exhibit trade-offs (negative spillovers) and obstruct the achievement of desired targets. This is, undoubtedly, the case of static models that build a network of interdependencies among development indicators. In some of these studies first order or second order effects are estimated \citep{leblanc_integration_2015,pradhan_systematic_2017,weitz_systemic_2018}. In others, different centrality measures (degree, eigenvector, betweenness and closeness) are calculated with the purpose of identifying influential policy issues \citep{allen_prioritising_2018,zhou_sustainable_2017}. 

Unfortunately, none of these studies takes into consideration that incoming positive spillovers not only imply reinforcing effects but also create side benefits that transform the incentive structure of the functionaries in charge of implementing the pertinent policies. \cite{castaneda_how_2018} show that functionaries' contributions to their corresponding policies decrease when they receive substantial spillovers from other policies.  Consequently, the potential benefits from investing in highly-central policy issues may be offset (or even reversed) by the negative incentives emerging at the receiving end of the spillovers. For this reason, it is indispensable to develop methods that allow balancing reinforcing effects and distorting incentives.

\subsubsection{Network estimation}

Another fundamental problem in measuring policy coherence has to do with the techniques used to build and calibrate the network of interdependencies among development indicators. The subjective/qualitative approach employed in some of the previous studies has important drawbacks: a discretionary emphasis on some portions of the network; high dependence on stakeholder and topical-expert knowledge; and erroneous interpretations derived form the wording of targets.\footnote{ For instance, \cite{leblanc_integration_2015} analyze this information to establish connections in a bipartite graph of targets and goals which, in turn, is used to project a network of goals where the weight of each edge is given by the number of targets linking different goals.} Besides the aforementioned shortcomings, these approaches are not scalable and, hence, their analyses are limited to a narrow policy space, even if there are more indicators available.\footnote{ Qualitative assessments are also popular in systems dynamics models, where the estimated network is combined with input-output data and a social accounting matrix \citep{pedercini_dynamic_2010}.}

When it comes to the quantitative estimation of links, their directions and weights are typically inferred through correlations \citep{pradhan_systematic_2017,zhou_sustainable_2017}, Bayesian techniques \citep{ceriani_multidimensional_2016,cinicioglu_exploring_2017,czyzewska_bayesian_2014}, or co-occurrence methods \citep{el-maghrabi_sustainable_2018}.\footnote{Co-occurrence methods, consist in determining, first, if the performance of a country in a particular indicator is above the average of countries with similar per capita income; then, by inferring the likely co-occurrence between two indicators (\emph{i.e.,} if they have proximate mechanisms for delivering above-average performances).} None of these approaches attempt to formally establish causal relationships, yet their results are interpreted as if the established links were causal connections. In Bayesian-network studies, however, some form of structural dependency is formulated. Clearly, the estimation of development-indicator networks would be greatly benefited from more data-driven techniques.

\subsubsection{Context specificity}

In the literature of economic development, it is well known that context matters for the success of a particular policy package \citep{rodrik_one_2009}. Several attempts have been made to quantify `context' in the literature of policy coherence. However, important drawbacks persist in all of them. For instance, interdependency networks are frequently estimated by pooling data from several countries. In some cases, the pooled sample consists of countries with radically different structures \citep{ceriani_multidimensional_2016,cinicioglu_exploring_2017,czyzewska_bayesian_2014,el-maghrabi_sustainable_2018}. In others, country-specific networks are derived (with a few tweaks) from a `master' network that was previously built for analyzing other countries \citep{pedercini_dynamic_2010}. A similar problem arises by not acknowledging country-specific political economy considerations such as governance factors (\textit{e.g.}, rule of law and monitoring corruption). The fact that these studies neglect such essential features seems contradictory since good policy planning must be aware of the policymaking process.

Accordingly, measurements of policy coherence that are developed with the aim of guiding real-world policies should consider country-specific spillover networks. When this is not possible due to data unavailability, spillover networks should be estimated with information from a reduced sample of countries with structural similarity, which can easily be obtained from clustering analysis. Furthermore, when possible, it is important to avoid parameters calibrated from analyses of other countries. Instead, it is more convenient to specify these parameters as social constructs (\textit{e.g.}, probability of catching a corrupt official), which can be derived from stable functional relationships between variables that capture alternative sources of country-specific information \citep{castaneda_evaluating_2018}.

\subsubsection{Implicit benchmark}

We have established the importance of context in the evaluation of policy coherence. Because of this, it is essential to consider the specific set of goals that a country wants to pursue. In other words, policy priorities can be entirely different depending on whether governments emphasize environmental issues, security concerns, inclusiveness goals, or a Scandinavian development model, for example.  We can argue, anew, that conceiving coherence just in terms of synergies and trade-offs provides an incomplete picture. For example, a policy issue like `access to electricity' may be central in a network because many other issues rely on this resource. However, if achieving a specific set of targets does not require transforming the electric grid, then network centrality becomes uninformative about the importance of this issue. Consequently, the policy advice spawned from this metric will generate an allocative inefficiencies through over-expenditure in this topic and under-expenditure in others. 

The relevance of context-specificity does not refer to unique patterns of interdependencies exclusively, but also to the presence of a particular government objective function. To the extent of our knowledge, we are not aware of any study that takes this idea into consideration.\footnote{Instead, one can commonly find vague expressions for the need for tools that can meet the 2030 Agenda for the SDGs.} In the end, governments' objective functions in developing countries are immensely influenced by multilateral organizations but always adapted to meet the countries' specific idiosyncrasies.

\subsection{A new definition}\label{sec:definition}

The non-observability of policy priorities, a variety of causal channels, the presence of spillover effects, the need for country-specificity, and the prerequisite to establish development targets in advance make extremely difficult to assess the coherence of policy prescriptions; even in quantitative analyses. In particular, combining tools that address these challenges is a daunting task on its own. 

We propose that counter-factual computational simulation can help overcoming these obstacles. In particular, we find agent-based models (ABMs) particularly well-suited for the task due to their flexibility to incorporate highly-detailed factors such as spillover networks and adaptive behavior. Hence, the ability to generate detailed counterfactuals leads us to think about policy coherence in terms of reference points. More formally, we define coherence of policy priorities as follows:

\begin{definition}
The policy priorities of country X are coherent with a set of targets T if the allocation of resources P destined to transformative policies is similar to the allocation Q that X would discover by trying to reach T.
\end{definition}

The previous definition is general enough to consider different policymaking processes. For example the process that generates $P$ and $Q$ could be the political economy game in PPI, but it could originate from alternative mechanisms as well. In a subtle way, this definition addresses the five challenges previously raised. First, the definition requires a pre-specified set of targets. These could be hypothetical values of development indicators, or the observed levels corresponding to a different nation $Y$ that $X$ would like to imitate (also called the `$Y$ development mode'). Second, it requires a retrospective estimation of the allocation profile $P$ through which the government destines resources to transformative policies (\textit{i.e.}, those that are not already committed, such as debt payments and infrastructure maintenance). Third, $Q$ is counterfactual in nature, which means that it needs to be estimated from a model where $X$ would set $T$ as its development targets and, then, try to reach them. Fourth, trying to reach targets involves a discovery process. This means that, in establishing its policy priorities, the government has to deal with country-specific factors such as inefficiencies and spillover effects. Note that our definition of coherence also allows for different distance metrics between $P$ and $Q$.

\section{Data and methods}\label{sec:methods}

\subsection{Data}

We use data on 79 development indicators at the country level. They come from three different sources: the World Economic Forum's Global Competitiveness Report, the World Development Indicators, and the World Governance Indicators; the latter two produced by the World Bank. The dataset consists of annual observations for 117 countries, covering the 2006–2016 period, the indicators have been normalized between 0 and 1 and they have been readjusted so that better outcomes translate into positive changes (see \cite{castaneda_how_2018} for more information). In order to provide initial summary statistics, we aggregate the indicators into 13 development pillars. In addition, we divide countries into three groups and a singleton. The first group consists of nations that joined the OECD prior to Mexico. Arguably, some of these early members are exemplary nations on which Mexico based its development goals for the last three decades. With exception of Iceland and Luxemburg, all OECD early members are in our dataset. The second group consists of countries with higher income per capita (IPC) than Mexico (it excludes those in group 1). These countries are useful for comparative purposes when describing the data. Group 3 is the singleton of Mexico. Finally, group 4 contains all countries with a lower IPC than Mexico.

\begin{figure}[h!]
    \centering
    \caption{Development pillars and income groups}
    \includegraphics[scale=.6]{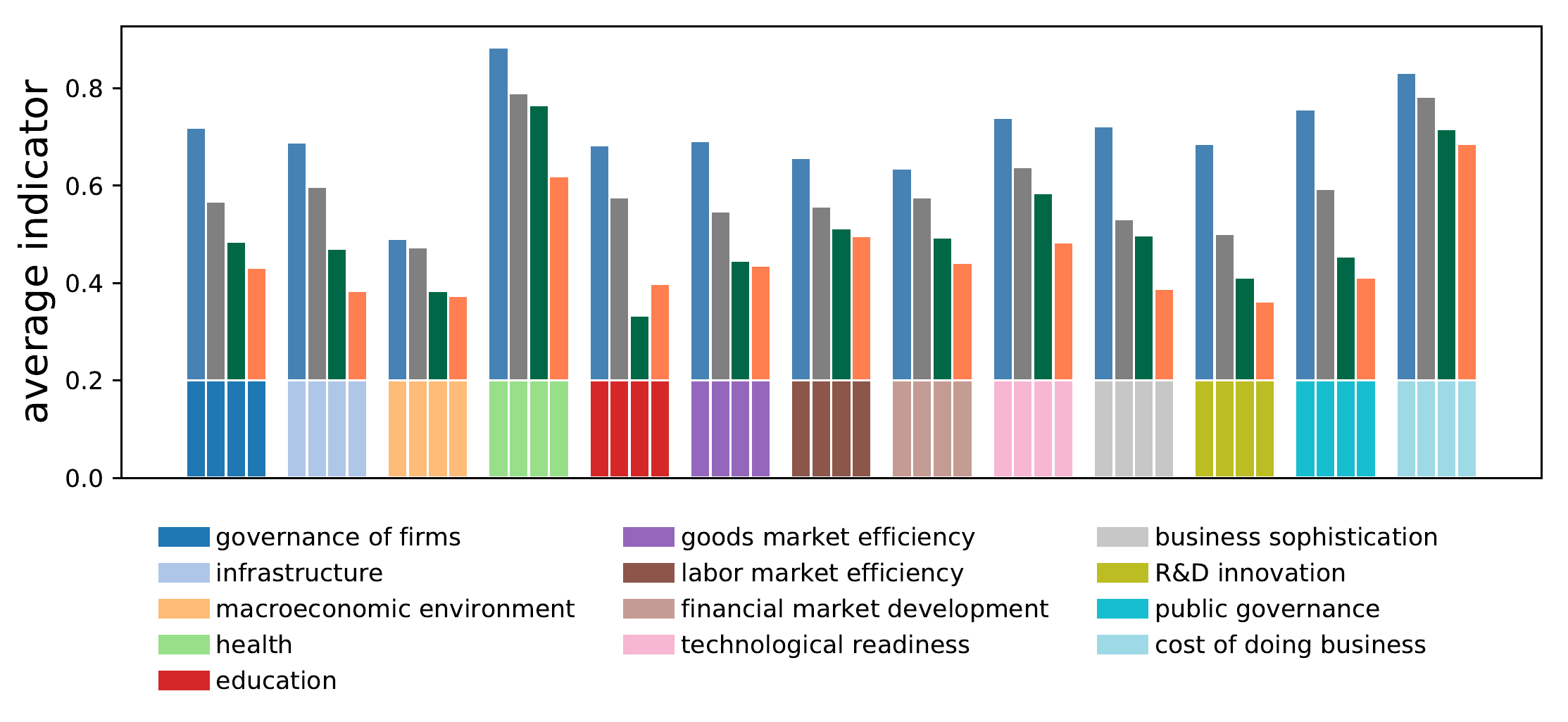}
    \caption*{\footnotesize{Average level of development indicators by pillar and group. The base color of each bar corresponds to a development pillar. Each bar within a pillar corresponds to each group. That is, the blue bars correspond to group 1 (OECD early members), the grey bars to group 2 (higher IPC than Mexico), the green bars to Mexico, and the orange bars to group 4 (lower IPC than Mexico).}}
    \centering\label{fig:indicators}
\end{figure}

Figure \ref{fig:indicators} displays the average level of development indicators of each group across the 13 development pillars (one color per group --bars-- and one color per pillar --bases). The first feature to highlight from these bars is that average levels vary significantly across pillars and across groups. Note that, in general, the more advanced a country is (left bars being more developed), the higher its average indicators. For the Mexican case (dark green bars), the largest differences with respect to the OECD early members (blue bars) correspond to the \textit{education} and \textit{public governance} pillars, while the smallest ones are in the \textit{macroeconomic environment}, \textit{cost of doing business} and \textit{health} pillars.

\subsection{Spillover networks}\label{sec:net}

In order to estimate the spillover network, we adopt an empirical strategy developed in the estimation of neural networks from functional magnetic resonance imaging data \citep{smith_network_2011,hoyer_causal_2008}. This strategy consists of two steps. First, we identify which pairs of indicators have a significant relationship (and their weights), through the method of triangulated maximally filtered graphs (TMFG) \citep{massara_network_2017}. The TMFG approach is a refinement of the planar maximal filtered graphs method \citep{tumminello_tool_2005}, which was first developed to identify influential assets in the US stock market \citep{kenett_dominating_2010}. Second, we infer the causal direction of these relationships through the method developed by \cite{hyvarinen_pairwise_2013}. This approach determines the direction of an edge by computing the likelihood ratio of two structural-equation models.

\begin{figure}[h!]
    \centering
    \caption{Spillover networks for different countries}
    \includegraphics[scale=.5]{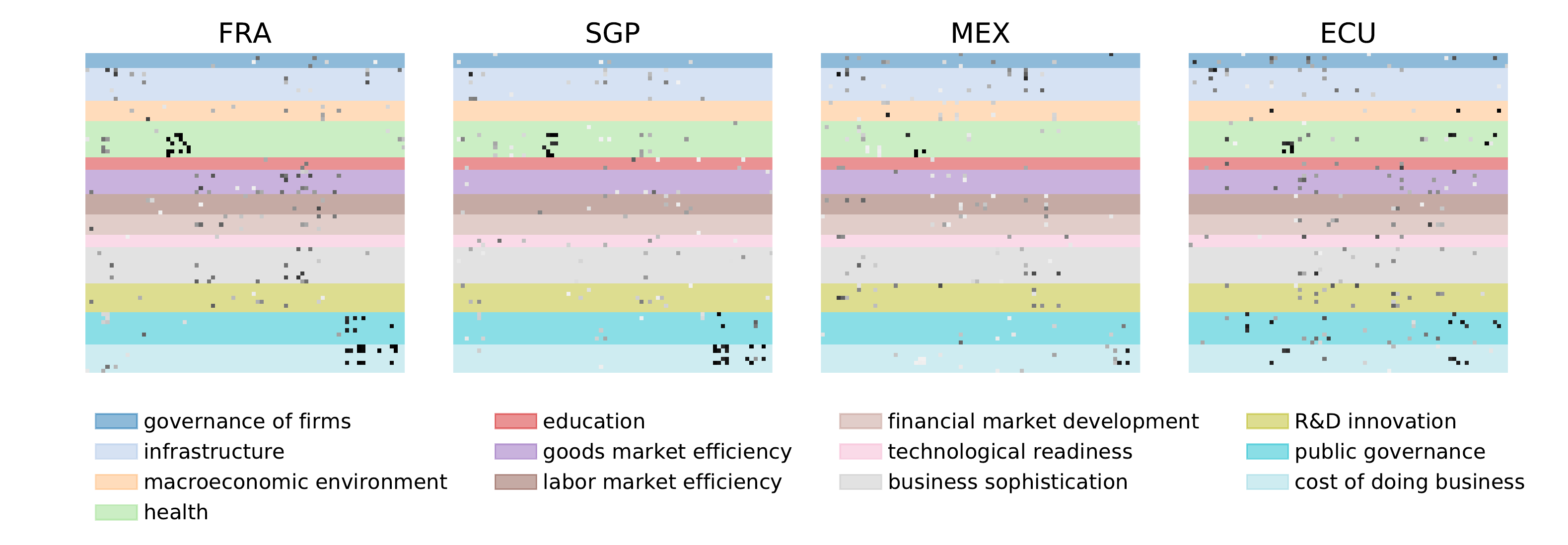}
    \caption*{\footnotesize{Each panel depicts the adjacency matrix of the spillover network of a country that belongs to each group: France (group 1), Singapore (group 2), Mexico (group 3) and Ecuador (group 4). The nodes (policy issues) have been arranged and colored according to the 13 development pillars. The gray-scale dots denote the presence and weight of edges (darker means more weight).}}
    \centering\label{fig:networks}
\end{figure}

Figure \ref{fig:networks} presents four spillover networks in terms of their adjacency matrices, one for each group. We present the 79 indicators in the same order for rows and columns. The colored segments indicate the pillars where the corresponding indicators are classified. The dots denote the presence of a link between two policy issues (nodes), and their weights are described with a gray-scale (darker means more weight). The fact that the dots above the diagonal are not a mirror image of the dots below it shows that the influence of many indicators runs in one specific direction (contrary to the co-occurrence methodology presented in \cite{el-maghrabi_sustainable_2018}). The dots that are distant from the diagonal indicate that there is a considerable number of connections between pillars. Note that the four panels present networks with remarkably different topological structures, confirming the relevance of a context-specific modeling approach. For example, France and Singapore (the two advanced nations in this example) have a structure where within-pillar links are plentiful, while Mexico and Ecuador exhibit much more off-diagonal edges; this feature reveals a more intricate structure, where many different policy issues seem to be interrelated.  

\subsection{Policy priority inference}

The framework of Policy Priority Inference was developed by \cite{castaneda_how_2018}, and has been previously used to estimate the resilience of development policies \citep{castaneda_resilience_2018}. The main idea of this approach is to estimate the policy priorities of governments by specifying a political economy game on a spillover network. In this game, the central authority allocates resources to different policy issues in order to achieve a given set of targets in its development indicators. The game takes place when the incentives of the officials in charge of the public policies are misaligned with the ones of the central authority, producing technical and allocative inefficiencies (for a further clarity, we provide the main features of the model in appendix \ref{app:ppi}).

Through PPI, we can estimate the policy priorities of any country in the sample. For this, we take a vector of development-indicator initial values (the initial conditions), final values\footnote{Here, we assume that the observed final values are informative of the targets that the government had in mind at that time. While this assumption is not ideal, we would argue that it suffers from significantly less distortions than using development indicators as policy priorities.} (the targets $T$) and a spillover network (estimated in section \ref{sec:net}). PPI simulates the evolution of the indicators from their initial values to the targets and, then, estimates the allocation profile $P$ that the country established during the sampling period. Furthermore, by defining a counterfactual set of targets, PPI can simulate a allocation profile $Q$ that works as a benchmark for the country's established priorities, which we use to construct our coherence index.

Formally, for $1,\dots,N$ indicators, PPI takes a vector $T$ of targets, a vector $I$ of initial indicators, and an adjacency matrix $\mathbb{A}$ of spillover effects as inputs. Among several outputs, we are interested in the vector $P$ of estimated allocations ($N$ allocations in total, one per indicator). Then, for an alternative set of targets, the output is vector $Q$ (see appendix \ref{app:calibration} for some technical details on model calibration).

\section{Analysis and results}\label{sec:results}

\subsection{Inferred policy priorities}

\begin{figure}[h!]
    \centering
    \caption{Estimated and na\"ive allocation profiles for Mexico}
    \includegraphics[scale=.4]{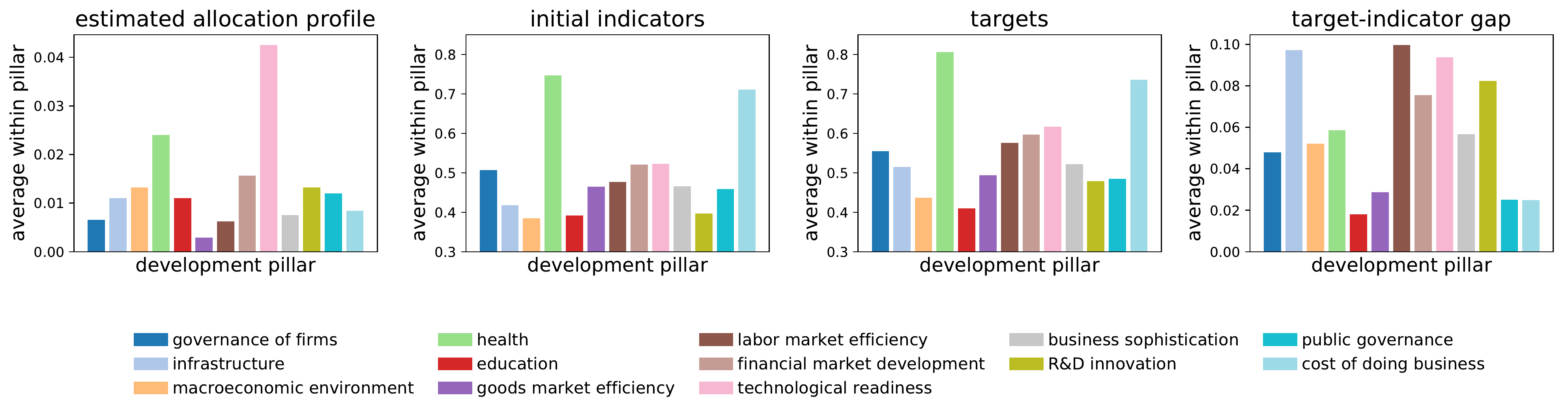}
\caption*{\footnotesize{All development-indicator data has been aggregated into 13 pillars. Panels from left to right: first, estimated allocation profile during the sampling period; second, initial development indicators; third, final indicators; fourth, differences between targets and initial indicators.}}
    \centering\label{fig:retrospective}
\end{figure}

First, we use the model's main outputs --the allocation profiles-- to demonstrate that naive observations from development indicators (or related data) are uninformative of policy priorities. Figure \ref{fig:retrospective} shows Mexico's allocation profile (left panel), aggregated into 13 development pillars. The remaining three panels show different forms of conceiving priorities with the same socioeconomic indicators used as inputs for PPI. These simplistic formulations are commonly used by development analysts and technocrats. Clearly, none of these three panels resemble the allocation profile estimated through PPI. For instance, the top pillar in the allocation profile, \textit{technological readiness} is not even the second in any of the other data configurations. This not only speaks of the non-triviality of the PPI estimations, but of the importance of generating policy evaluation through theoretically-founded models of the policymaking process. Evidently, the also-common practice of evaluating coherence according to speeches and official documents suffers from similar pitfalls.

\subsection{Counterfactuals as consistent priorities}

We have argued that the coherence of policy priorities $P$ is relative to context-specific considerations and to the specification of targets $T$. We have also stated that the simulated allocation profile $Q$ is the result of discovering priorities by trying to achieve $T$. That is, $Q$ represents what a country would do if it was \textit{really} committed to reach $T$, while being constrained by the political economy and the spillover network. Hence, let us call $Q$ the consistent allocation profile.

Consistent profiles are specific to the targets that a country wants to pursue. In the literature of development economics, it is common to think about $T$ as the development indicators of an exemplary nation. This is the basic principle behind Akamatsu's flying geese \citep{akamatsu_historical_1962} and the idea of `development footprints'. Here, we formalize this view through the concept of development modes. To illustrate, if Mexico's government determines that its development goals should be France's current indicators, we say that Mexico adopts the France development mode.\footnote{Of course, we could think of mixtures of exemplary countries or configurations of $T$. However, exploring the vast space of mixtures of development modes is an endeavour that is out of the scope of this paper. To get our point across, it suffices to analyze coherence in the context of early OECD member countries as development modes.} 

For each development mode, a country --say Mexico-- has a consistent allocation profile. For a given development mode, we estimate this profile by setting Mexico's target to be the initial (2006) values of the development mode in question. In other words, the estimated $Q$ represents the policy priorities that Mexico would follow if it was serious about catching up another nation (the development mode). The intuition is straightforward: development-indicator data is not informative about the priorities of a government, but rather of an observable and achievable state to which another nation can aspire.

\begin{figure}[h!]
    \centering
    \caption{Retrospective and consistent allocation profiles for Mexico}
    \includegraphics[scale=.4]{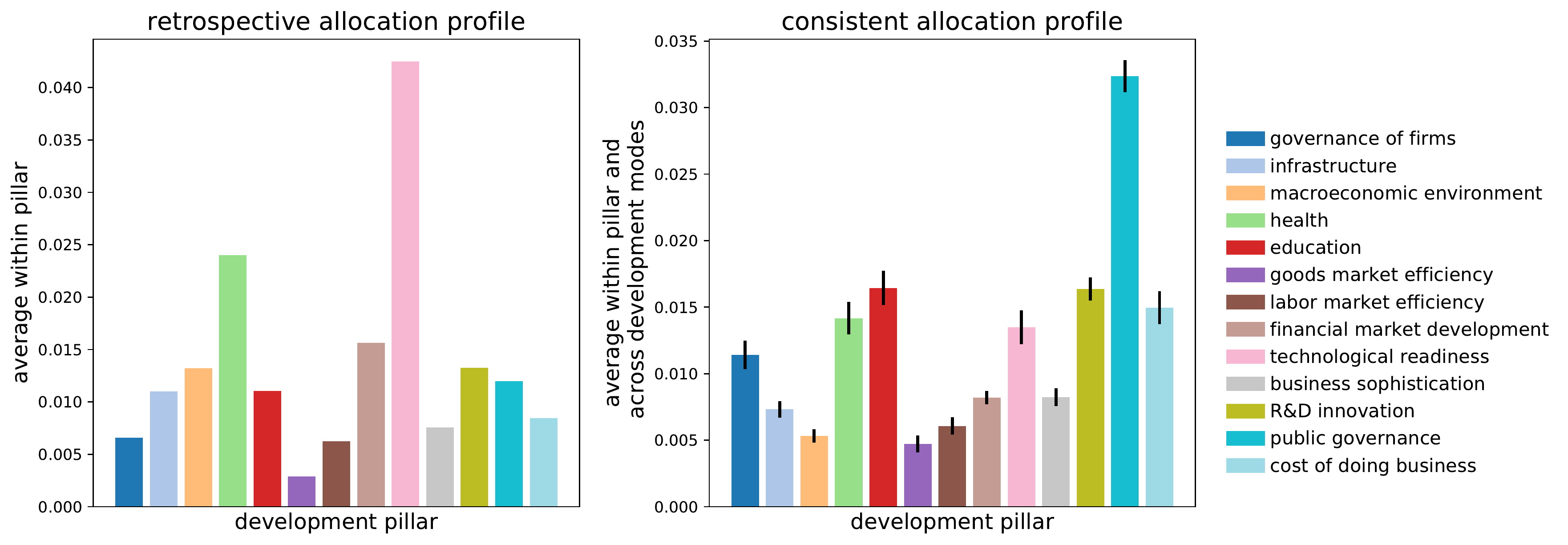}
\caption*{\footnotesize{All development-indicator data has been aggregated into 13 pillars. The vertical black lines in the right panel denote the standard errors from the cross-national aggregation of consistent allocation profiles.}}
    \centering\label{fig:consistent}
\end{figure}

Consistent allocation profiles may or not differ from those obtained through a retrospective estimation. This depends on whether the observed (final values) indicators of a country lead to a development path similar to that of the counterfactual targets. Figure \ref{fig:consistent} shows both the retrospective and the consistent allocation profiles for Mexico. In the left panel, we show the retrospective estimation. The right panel shows the average consistent allocation profile that was estimated for the 22 counterfactuals of adopting the development modes corresponding to the early OECD members. Clearly, in this case, the priorities suggested by both types of profiles are very different. For example, the right panel suggests that Mexico would have to invest heavily in \textit{public governance} if it wants to catch up with the OECD early members, while currently, it prioritizes \textit{technological readiness}. Notice that the top-five priorities for Mexico in the consistent profile, at the pillar level, are the following: \emph{public governance}, \emph{education}, \emph{R\&D innovation}, \emph{cost of doing business} and \emph{health}. For this reason, in this paper, we exploit the discrepancy between retrospective and consistent profiles to build a metric for the coherence of policy priorities.

\subsection{A coherence index}

\begin{figure}[h!]
    \centering
    \caption{Mexico's coherence: retrospective \textit{versus} consistent allocations for Mexico}
    \includegraphics[scale=.35]{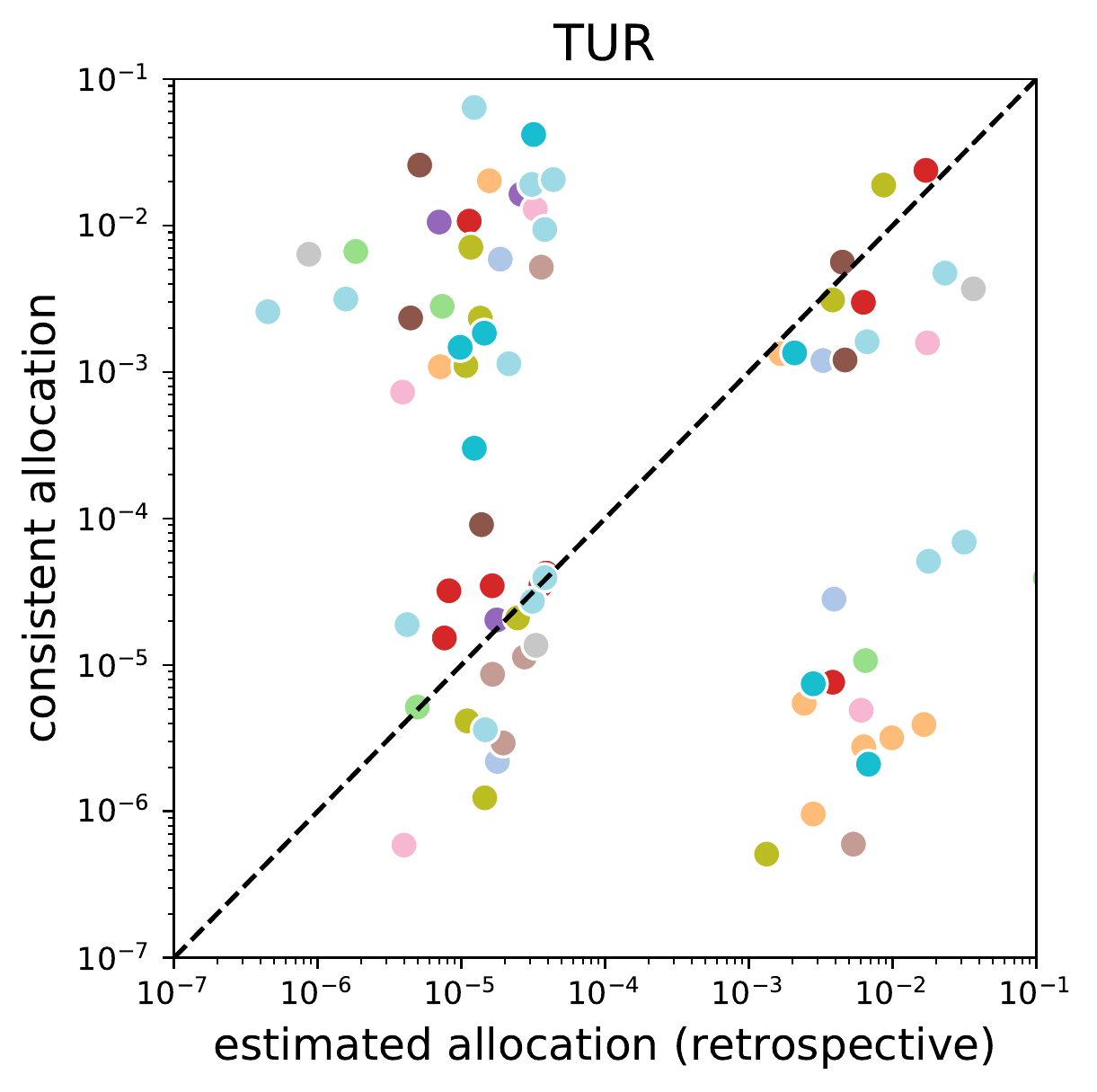}
    \includegraphics[scale=.35]{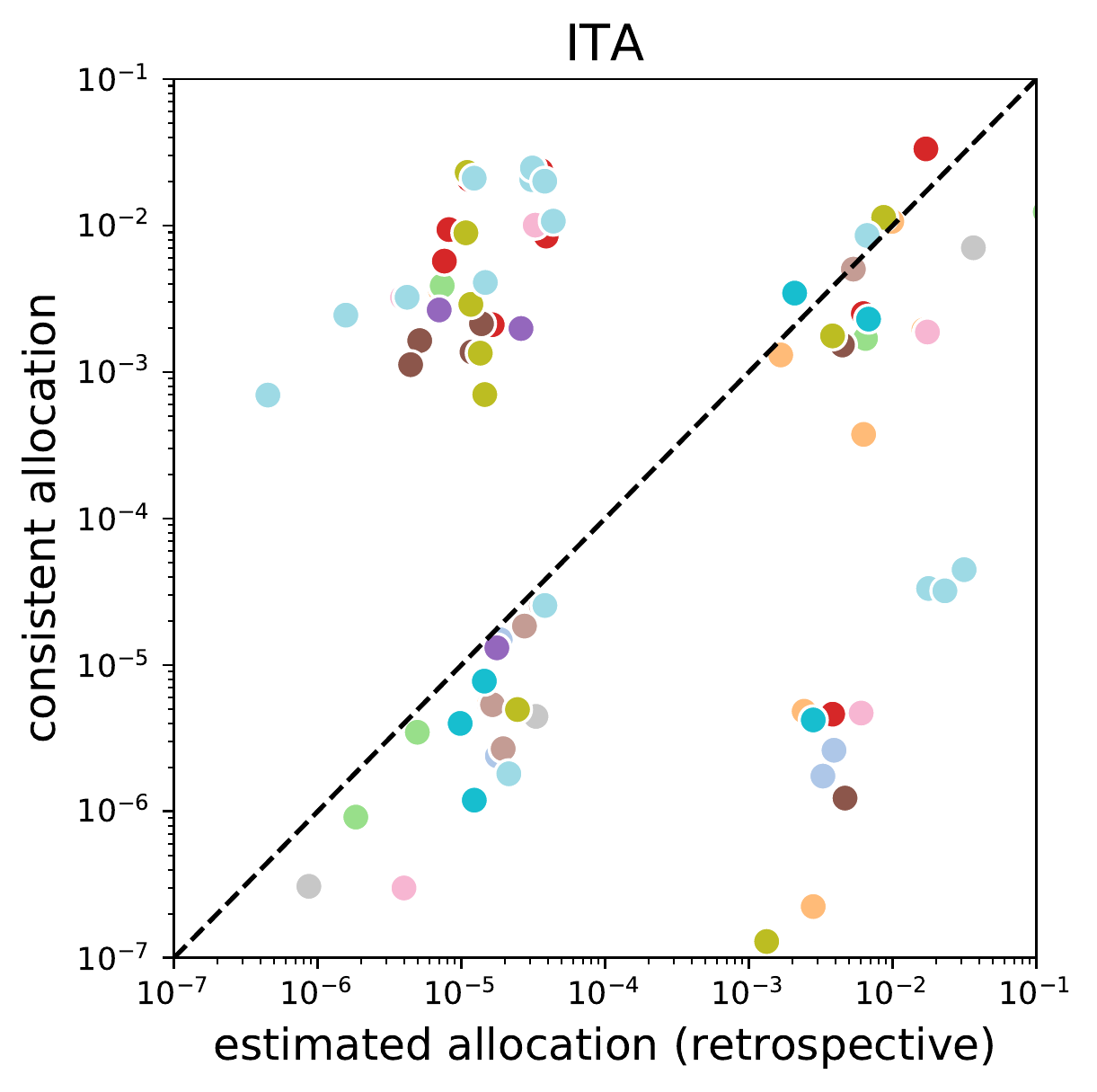}
    \includegraphics[scale=.35]{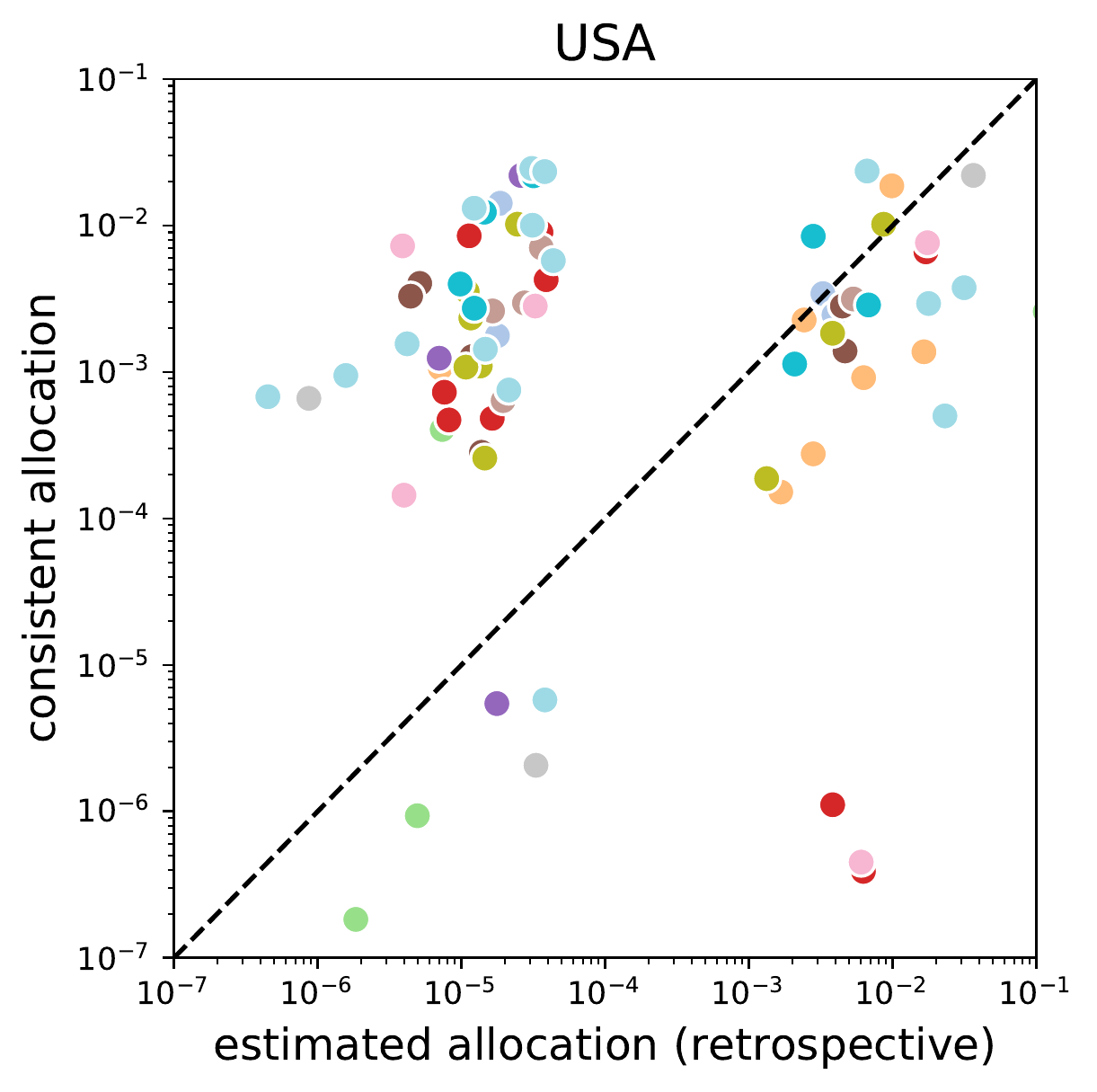}
\caption*{\footnotesize{Three examples of development modes that Mexico can adopt from the early OECD members. Full coherence implies that each dot representing a particular policy issue should be along the 45-degree line}}
    \centering\label{fig:coherence_cases}
\end{figure}

Our definition of coherence captures the disparity between the current policy priorities of a country and those that it would establish should it decided to adopt certain goals. This is achieved by comparing the retrospective allocation profile $P$ against the consistent profile $Q$. Let us illustrate how these two profiles differ across three development modes that Mexico could adopt from the OECD. Figure \ref{fig:coherence_cases} shows the indicator-level retrospective allocations (horizontal axes) and the consistent ones (vertical axes). If Mexico was coherent with each mode, the dots in the three panels would lie on the 45-degree line. A dot above the diagonal means that Mexico is under-spending in a policy issue that would receive more investment if the government would \textit{really} want to adopt that development mode. On the other hand, dots below the diagonal are policy issues where the government over-spends. In other words, these diagrams offer an insight of Mexico's allocative inefficiencies.

Figure \ref{fig:coherence_cases} gives us a good idea of the principle behind a metric for coherence: measuring discrepancies between $P$ and $Q$. To achieve this, we also need to consider the other side of the coin: \textit{policy incoherence}. Thus, a comprehensive metric should be informative about how well and how bad a country is doing. Thus, in this context, we construct an inconsistent allocation profile $R$ containing the exact opposite priorities as $Q$. In other words, the top priority in $Q$ is the lowest one in $R$, the second highest in $Q$ is the second lowest in $R$ and so forth. Hence, if the retrospective profile $P$ is very similar to $R$, we can say that the policies of the associated country are incoherent.

Next, we combine $Q$ and $R$ to construct the coherence index. Let us consider a metric $d(X, Y)$ measuring the distance between two allocation profiles $X$ and $Y$. Then, our coherence index is defined as

\begin{equation}
    h = \frac{d(P,R) - d(P,Q)}{d(P,R) + d(P,Q)}.
    \label{eq:index}
\end{equation}

Index $h$ works like a correlation coefficient, ranging from -1 to 1. If the index is negative, it means that the country is incoherent with respect to the given targets. If it is zero, it means that coherence/incoherence is ambiguous because the policy priorities are equally similar to the consistent and the inconsistent profiles. When $h$ is positive, we say that the policy priorities are coherent, and when $h=1$, we speak of full coherence. 

In a nutshell, this index provides a standardized metric to evaluate the degree of coherence/incoherence across countries and targets. Because targets $T$ are usually measured through development indicators, equation \ref{eq:index} also implies that, when the final indicators of a country are extremely similar to the initial ones of a development mode, $h \approx 1$. In these situations, development-indicator data can be informative about coherence. This, however, is an extremely rare case since, by definition, developing countries exhibit laggard indicators with respect to the developed nations. Furthermore, even if the ordering of indicators is the same between a country and its development mode, their distance increases the possibility of producing allocation profiles that do not follow such ordering.\footnote{In contrast, it is technically possible that different orderings and levels of targets yield a high coherence estimation. This scenario would happen if the lower level targets were in the development path of the high level targets.} In section \ref{sec:validation}, we present two validation cases and show, in appendix \ref{app:triviality}, that higher similarity (Pearson correlation) between indicators does not associate to more coherence.

In order to disambiguate an index, we can perform statistical significance tests. For this, we use the Monte Carlo simulations from PPI to obtain the distribution of $h$. Thus, the estimated index should be the expected value from this distribution, and its significance with respect to $h=0$ should follow from the chosen percentile of the corresponding distribution. Finally, our index can take different distance metrics (normalized for different amounts of indicators and weights). The qualitative nature of our empirical results is robust across several of them. Here, we present the outcome of using the simple metric: $d = \sum_i^N|X_i-Y_i|$.

\subsection{Results for Mexico}

Our main finding is that Mexico' policies are not coherent during the sampling period. Figure \ref{fig:mexico} shows the coherence indices estimated for the 22 possible development modes coming from OECD early members. From this, we can see that the index is negative in all cases. In addition, there is no association between the level of development of the country to imitate (here measured in income per capita) and Mexico's coherence. This suggests that, besides not being coherent, Mexico's government did not make a systematic effort to follow the most developed countries. In fact, within Mexico's different levels of incoherence, it seems that the Italy development mode yields the closest priorities to the Mexican retrospective ones. Furthermore, other less developed countries from the OECD such as Greece and Turkey are not associated to the lowest levels of incoherence, something desirable for a country that aspires to catch up with the developed world.

\begin{figure}[h!]
    \centering
    \caption{Mexico and its coherence with OECD development modes}
    \includegraphics[scale=.5]{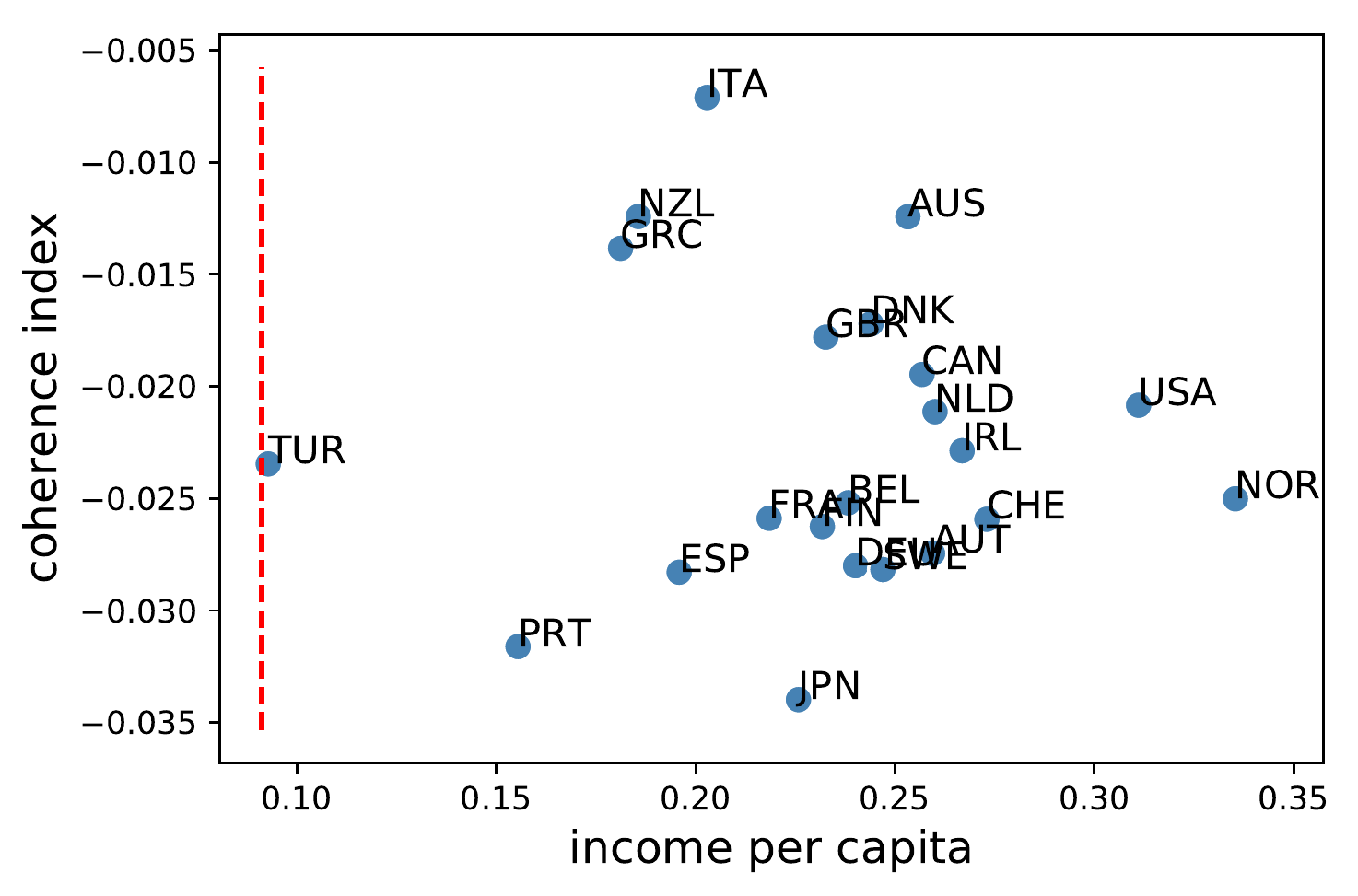}
\caption*{\footnotesize{The dashed red line denotes the income per capita of Mexico.}}
    \centering\label{fig:mexico_indicators}
\end{figure}

A virtue of PPI is its ability to provide disaggregated statistics of the data underlying the index. For instance, consider the disparities between $P$ and $Q$ at the level of each development indicator. This information can be extremely helpful to diagnose the sources of incoherence. Figure \ref{fig:mexico} shows the differences $P_i - Q_i$ for the best (Italy) and worst (Japan) development modes in terms of Mexico's coherence. This difference is presented at the level of each indicator $i$, providing a detailed picture of allocative inefficiencies. Here, a negative difference means that the Mexican government is under-spending with respect to the consistent allocation in $Q$, while the opposite sign denotes over-expenditure. As shown in the two panels, Mexico has been under-spending in most policy issues, yet it exercises a disproportionate over-expenditure in the policy issue of \textit{redundancy costs} (a component of the \textit{labor market efficiency} pillar). Other two important topics where government over-spends are \textit{tuberculosis cases} and \textit{general government debt}. Apparently, the burden of \textit{public debt}, \textit{transaction costs in the labor market} and expenses in certain diseases have become a hurdle for the implementation of more coherent policies. These results do not mean that excessive expenditures have to be canceled, it rather implies that the problems associated to these issues have to be solved. 

Although both development modes exhibit a similar pattern of allocative inefficiencies, we can pinpoint some differences. For instance, there are more cases of marginal over-expenditures in the Italian mode (\textit{e.g.}, in \textit{R\&D and innovation}). In the Japanese mode, under-spending in \textit{availability of the latest technology} and in \textit{extent of staff training} is more prevalent than in the Italian one. These differences show that attempting to reach Japanese targets require a more technological focus and better human capital. This exercise illustrates that the distribution of inefficiencies across policy issues depends on the type of development mode that a country wants to pursue. In contrast, technical inefficiencies (\textit{e.g.}, corruption) depend on the strength of governance indicators and the topology of network spillovers.

In principle, some of the resources required to foster under-spent policy issues could come from unnecessary allocations in other topics. Nevertheless, from Figure \ref{fig:mexico}, it is clear that policy issues with an under-expenditure are numerous in both examples. For example, investment in pillars like \textit{public governance}, \textit{R\&D and innovation}, \textit{health} and \textit{costs of doing business} (principally in \textit{crime and violence}) are especially large for the Italian and Japanese modes. Therefore, a major effort should be directed by Mexicans to create the budgetary leeway for financing these expenses and, as stated above, this starts by reducing the debt problem.

\begin{figure}[h!]
    \centering
    \caption{Mexico's allocative inefficiencies for two development modes}
    \includegraphics[scale=.45]{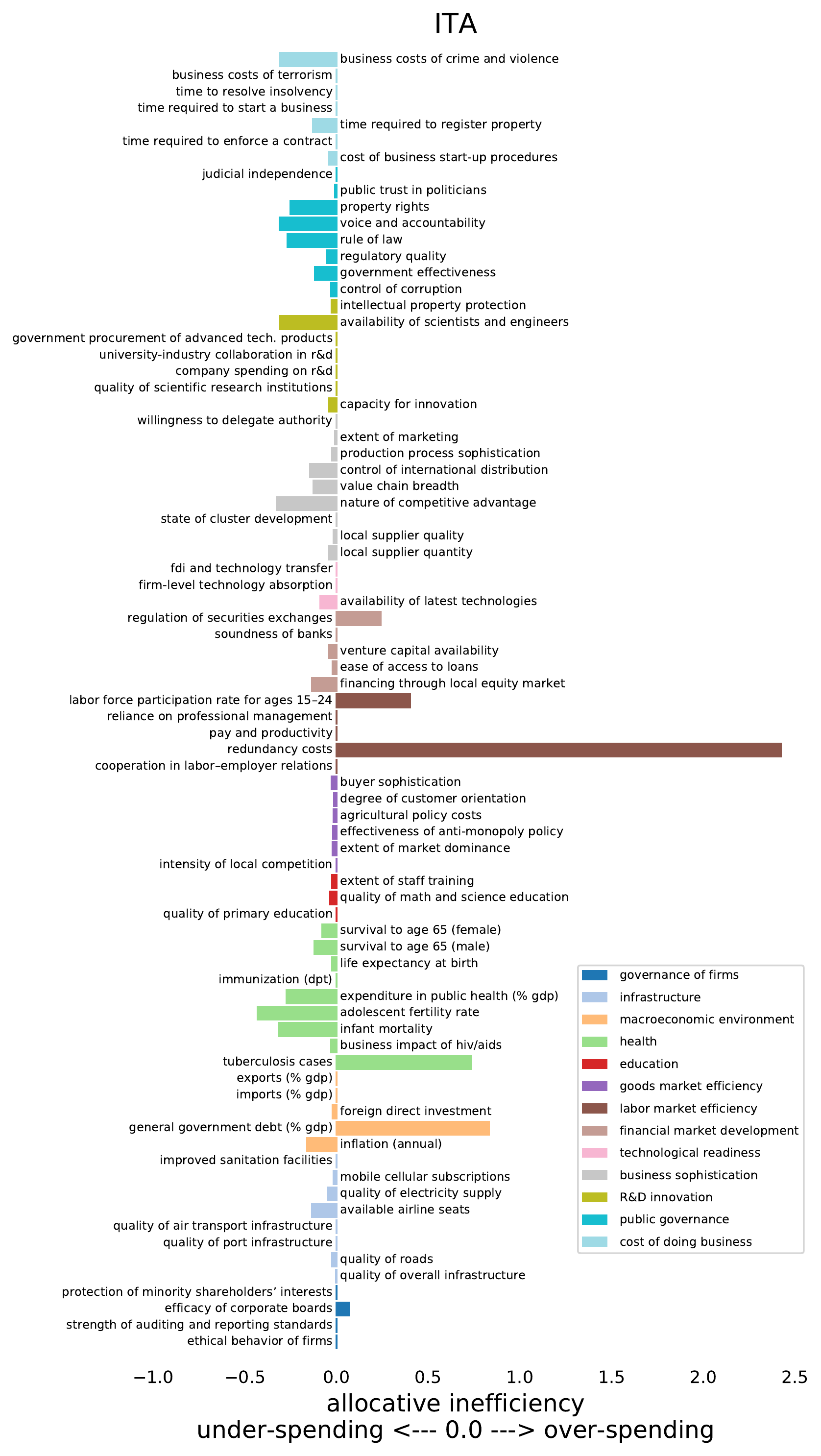}
    \includegraphics[scale=.45]{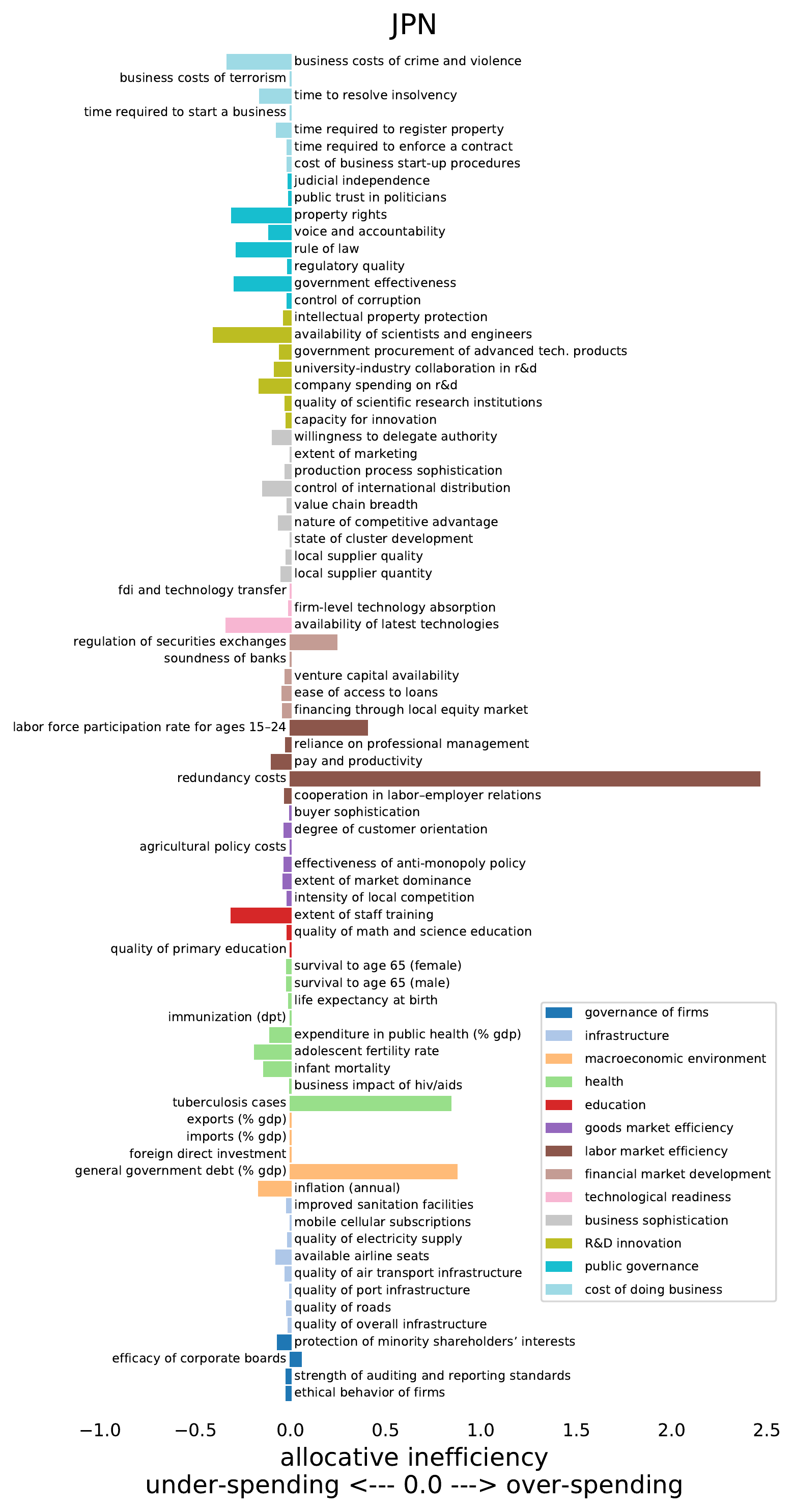}
\caption*{\footnotesize{The units in the horizontal axis have been re-scaled for clarity of exposition. Each color denotes a development pillar. Positive bars indicate over-expenditure, while negative bars under-expenditure.}}
    \centering\label{fig:mexico}
\end{figure}

\subsection{Validation}\label{sec:validation}

In this section, we perform `soft' validation tests by looking for consistencies between out empirical results and well known development experiences from other countries.\footnote{\cite{castaneda_how_2018} provide rigurious external and external validation tests for PPI.} For this, we analyze another OECD member country that presents two conditions: $i$) it has had a clear development mode for several years, and $ii$) it has been successful in achieving economic development. Hence, our country of choice is South Korea, who became an OECD member in 1996, two years after Mexico. Presumably, we can apply a similar logic as in the Mexican case in terms of the incentives that the Korean government had in order to join this organization \citep{lee_economic_1996,hsu_lessons_2014}. Therefore, a validation should follow from $a$) Korea having predominantly positive and significant coherence indices and $b$) Japan being among the development modes with which Korea is most coherent.

\begin{figure}[h!]
    \centering
    \caption{Two validation cases}
    \includegraphics[scale=.45]{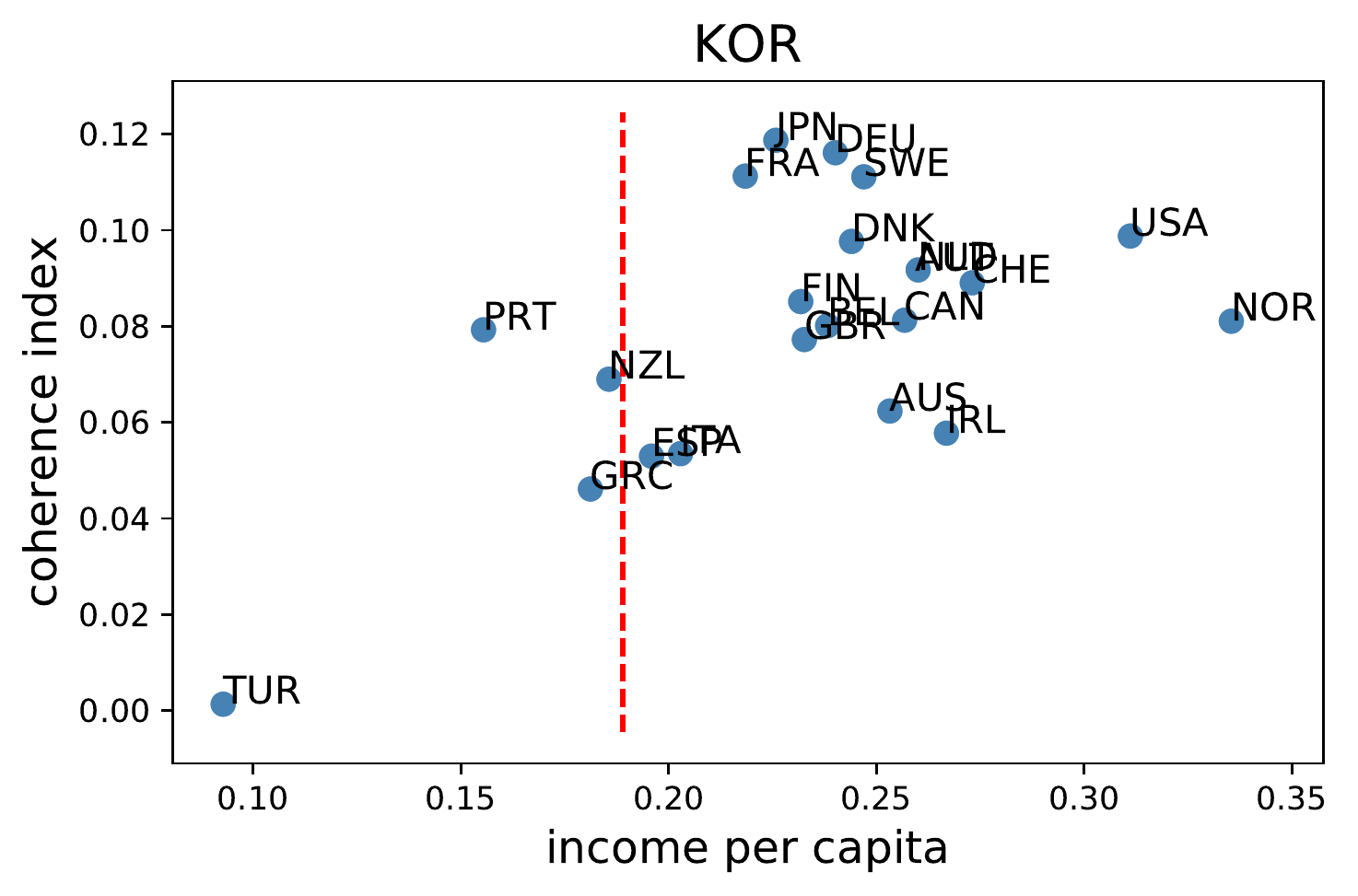}
    \includegraphics[scale=.45]{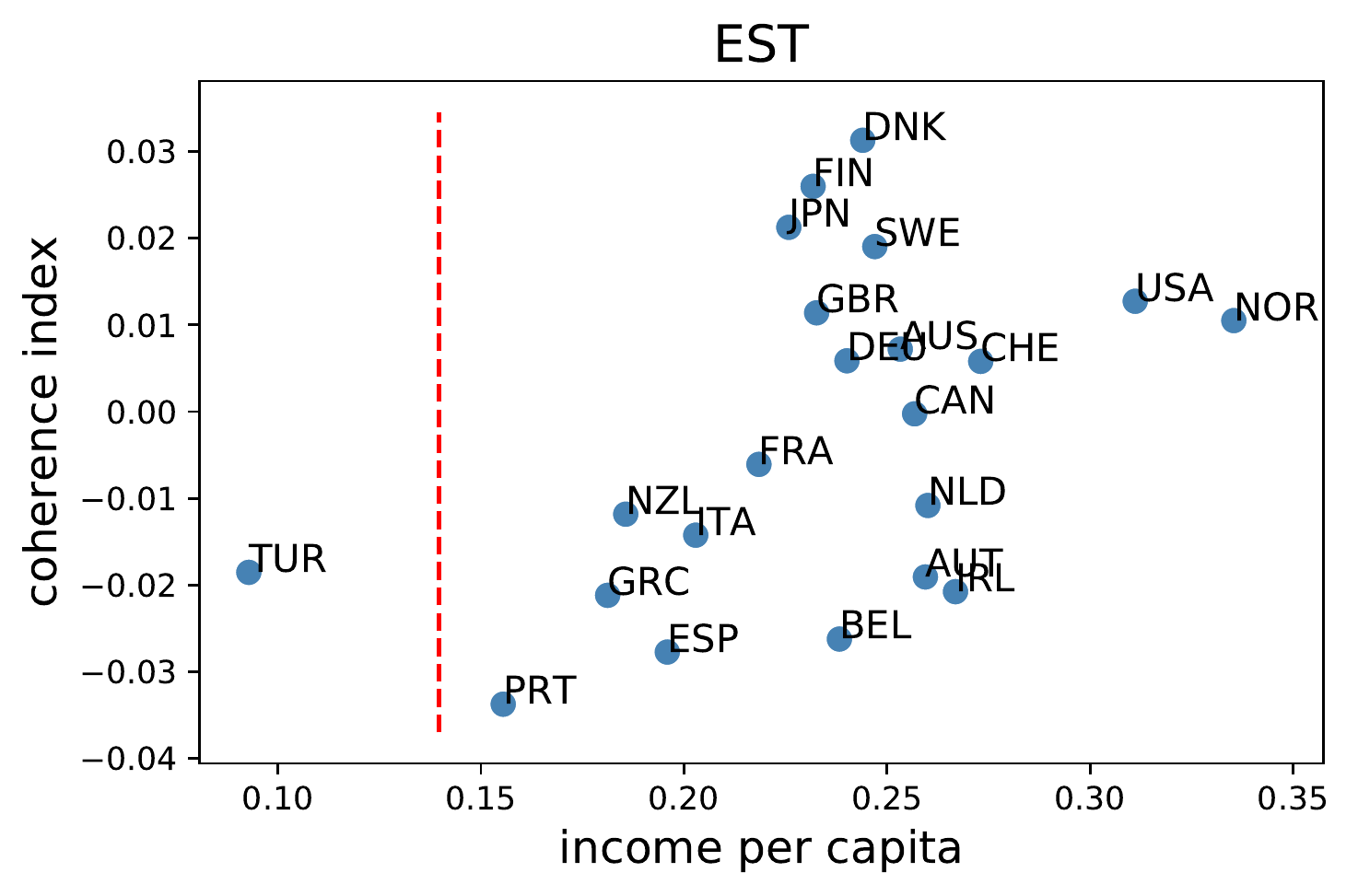}
\caption*{\footnotesize{The dashed red line denotes the income per capita of the developing country under study.}}
    \centering\label{fig:validation}
\end{figure}

The left panel in Figure \ref{fig:validation} confirms the validity of out method. First, note that all the indices are positive. In fact, Korea has unambiguous indices in most cases (see Table \ref{tab:estimates} for estimates). Second, Japan has a prominent position since it is the development mode with which Korea is most coherent. This is consistent with well known studies of the economic development of Korea: ``\textit{Japan's development strategies have served as a model for Korean policy-makers}'' \citep[p.~13]{lee_sustaining_2007}. Moreover, other highly coherent modes include Germany and Sweden, which speaks of the visible transformation of Korea towards an outward-looking and technologically-oriented economy.

As an additional validation, we also present the case of Estonia, a Baltic country that is known to be undergoing important structural transformations. In contrast with Mexico and Korea, Estonia became an OECD member in 2010, half way through our sampling period. Here, our argument is that, during this time, Estonia was going through a process of aligning its priorities to the requirements of the OECD. Hence, one would expect less coherence than Korea. However, the Estonian case is interesting because, after its separation from the USSR, it has made serious attempts to adopt a Nordic development model \citep{virkus_internationalisation_2002,alestalo_nordic_2009}. Thus, further validity of our method should be reflected in higher indices for the Nordic development modes. This is confirmed in the right panel of Figure \ref{fig:validation}, where Denmark, Finland and Sweden are 3 of the 4 countries with highest coherence. Moreover, the other top country is Japan, which is consistent with the impressive technological transition that Estonia is currently experiencing \citep{sinani_spillovers_2004,tiits_technologyintensive_2007}.\footnote{It is also intriguing that there is a very strong correlation between our coherence index for Estonia and the Inglehart-Welzel index for traditional secular/rational values in the 2010-2014 World Value Survey \citep{inglehart_world_2014}. Perhaps, it is not so adventurous to hypothesize that a country's true development goals are largely defined on cultural grounds.}

\subsection{Results for other OECD members}

So far, we have shown that our index suggests a lack of coherence in the Mexican policy priorities, while being congruent with the expected outcomes of two country cases. However, there are other 9 developing countries that became members after 1994, so the reader might be interested in comparing their coherence indices. This, of course, with the caveat that their reasons for joining the OECD are not as clear as for Mexico, Korea and Estonia. Table \ref{tab:estimates} shows all the estimates of $h$ for each country-mode pair (highlighting the most coherent mode for each country). In contrast with a predominantly coherent Korea, Slovakia is the most incoherent nation among the OECD late members (all but three counterfactual targets have a negative and statistically significant value). In addition, note that a country like the Czech Republic can be relatively coherent with one development mode (Sweden) while, at the same time, be incoherent with another (Turkey).

\begin{table}[h!]
    \caption{Coherence indices for augmented sample}
    \footnotesize
    \begin{center}
        \begin{tabular}{ c c c c c c c c c c c c c c}
            \hline
            Mode & CHL & CZE & EST & HUN & ISR & KOR & LTU & LVA & MEX & POL & SVK & SVN\\
            \hline
AUS&$0.1^{*}$&$0.03^{}$&$0.01^{}$&$-0.01^{}$&$-0.04^{}$&$0.06^{}$&$0.03^{}$&$-0.03^{}$&$-0.01^{}$&$0.1^{*}$&$-0.05^{**}$&$0.05^{}$\\
AUT&$0.07^{}$&$0.07^{}$&$-0.02^{}$&$-0.02^{}$&$-0.04^{}$&$0.09^{**}$&$0.01^{}$&$-0.01^{}$&$-0.03^{}$&$0.1^{*}$&$-0.04^{*}$&$0.06^{}$\\
BEL&$0.04^{}$&$0.1^{*}$&$-0.03^{}$&$0.02^{}$&$-0.03^{}$&$0.08^{*}$&$0.02^{}$&$-0.02^{}$&$-0.03^{}$&$0.11^{*}$&$-0.05^{**}$&$0.06^{}$\\
CAN&$0.09^{*}$&$0.04^{}$&$-0.0^{}$&$-0.0^{}$&$-0.04^{}$&$0.08^{*}$&$0.01^{}$&$-0.01^{}$&$-0.02^{}$&$0.1^{*}$&$-0.04^{}$&$0.06^{}$\\
CHE&$0.08^{*}$&$0.09^{*}$&$0.01^{}$&$0.0^{}$&$-0.02^{}$&$0.09^{*}$&$0.02^{}$&$-0.01^{}$&$-0.03^{}$&$0.09^{*}$&$-0.03^{}$&$0.06^{}$\\
DEU&$0.09^{*}$&$0.1^{**}$&$0.01^{}$&$0.01^{}$&$-0.03^{}$&$0.12^{**}$&$0.01^{}$&$-0.01^{}$&$-0.03^{}$&$0.11^{*}$&$-0.05^{**}$&$0.06^{}$\\
DNK&$0.07^{}$&$0.1^{**}$&\cellcolor{blue!25}$0.03^{}$&$-0.0^{}$&$-0.02^{}$&$0.1^{**}$&$0.04^{}$&$-0.02^{}$&$-0.02^{}$&$0.09^{*}$&$-0.04^{}$&$0.06^{}$\\
ESP&$0.01^{}$&$0.01^{}$&$-0.03^{}$&$-0.04^{}$&$-0.04^{*}$&$0.05^{}$&$0.01^{}$&$-0.03^{}$&$-0.03^{}$&$0.1^{*}$&$-0.04^{*}$&$0.02^{}$\\
FIN&$0.08^{*}$&\cellcolor{blue!25}$0.11^{**}$&$0.03^{}$&$0.02^{}$&$-0.02^{}$&$0.09^{*}$&$0.04^{}$&$-0.01^{}$&$-0.03^{}$&$0.1^{*}$&$-0.04^{*}$&$0.06^{}$\\
FRA&$0.07^{}$&$0.09^{*}$&$-0.01^{}$&$0.0^{}$&$-0.02^{}$&$0.11^{**}$&$-0.0^{}$&$-0.03^{}$&$-0.03^{}$&$0.11^{*}$&$-0.05^{**}$&$0.05^{}$\\
GBR&\cellcolor{blue!25}$0.11^{**}$&$0.09^{*}$&$0.01^{}$&$-0.01^{}$&$-0.03^{}$&$0.08^{*}$&$0.02^{}$&$-0.03^{}$&$-0.02^{}$&\cellcolor{blue!25}$0.11^{**}$&$-0.04^{*}$&$0.06^{}$\\
GRC&$-0.01^{}$&$0.0^{}$&$-0.02^{}$&\cellcolor{blue!25}$0.07^{*}$&$-0.01^{}$&$0.05^{}$&$-0.01^{}$&$0.03^{}$&$-0.01^{}$&$0.07^{}$&$0.0^{}$&$0.02^{}$\\
IRL&$0.07^{}$&$0.06^{}$&$-0.02^{}$&$0.0^{}$&$-0.04^{}$&$0.06^{}$&$0.01^{}$&$-0.03^{}$&$-0.02^{}$&$0.1^{*}$&$-0.05^{**}$&$0.04^{}$\\
ITA&$0.01^{}$&$0.0^{}$&$-0.01^{}$&$0.02^{}$&$-0.01^{}$&$0.05^{}$&\cellcolor{blue!25}$0.07^{}$&\cellcolor{blue!25}$0.05^{}$&\cellcolor{blue!25}$-0.01^{}$&$0.06^{}$&\cellcolor{blue!25}$0.03^{}$&$0.03^{}$\\
JPN&$0.06^{}$&$0.08^{*}$&$0.02^{}$&$0.01^{}$&\cellcolor{blue!25}$-0.0^{}$&\cellcolor{blue!25}$0.12^{**}$&$0.05^{}$&$0.01^{}$&$-0.03^{*}$&$0.09^{}$&$-0.02^{}$&$0.06^{}$\\
NLD&$0.08^{}$&$0.08^{*}$&$-0.01^{}$&$-0.0^{}$&$-0.03^{}$&$0.09^{*}$&$0.02^{}$&$-0.01^{}$&$-0.02^{}$&$0.1^{**}$&$-0.04^{*}$&$0.06^{}$\\
NOR&$0.1^{*}$&$0.06^{}$&$0.01^{}$&$-0.01^{}$&$-0.03^{}$&$0.08^{*}$&$0.02^{}$&$-0.01^{}$&$-0.03^{}$&$0.09^{*}$&$-0.04^{*}$&$0.06^{}$\\
NZL&$0.08^{}$&$0.04^{}$&$-0.01^{}$&$0.0^{}$&$-0.02^{}$&$0.07^{}$&$0.04^{}$&$-0.02^{}$&$-0.01^{}$&$0.1^{*}$&$-0.04^{}$&$0.05^{}$\\
PRT&$0.03^{}$&$0.02^{}$&$-0.03^{}$&$0.03^{}$&$-0.01^{}$&$0.08^{*}$&$0.0^{}$&$-0.01^{}$&$-0.03^{**}$&$0.08^{*}$&$-0.04^{}$&$0.02^{}$\\
SWE&$0.1^{*}$&$0.09^{**}$&$0.02^{}$&$0.0^{}$&$-0.02^{}$&$0.11^{**}$&$0.03^{}$&$-0.02^{}$&$-0.03^{}$&$0.1^{*}$&$-0.04^{**}$&$0.06^{}$\\
TUR&$0.01^{}$&$-0.03^{*}$&$-0.02^{}$&$-0.03^{}$&$-0.02^{}$&$0.0^{}$&$0.04^{}$&$-0.03^{}$&$-0.02^{}$&$0.05^{}$&$-0.01^{}$&$0.01^{}$\\
USA&$0.1^{*}$&$0.1^{**}$&$0.01^{}$&$-0.02^{}$&$-0.01^{}$&$0.1^{**}$&$0.04^{}$&$-0.02^{}$&$-0.02^{}$&$0.09^{*}$&$-0.04^{*}$&\cellcolor{blue!25}$0.07^{}$\\
\hline
Average&$0.07$&$0.06$&$-0.0$&$0.0$&$-0.02$&$0.08$&$0.02$&$-0.01$&$-0.02$&$0.09$&$-0.03$&$0.0$ \\
            \hline
        \end{tabular}
    \end{center}\label{tab:estimates}
    \caption*{\footnotesize{Shaded areas are the development modes with maximum coherence for each country. Stars indicate non-ambiguity significance. *:$p<0.1$, **:$p<0.05$}}
\end{table}

Overall, Table \ref{tab:estimates} suggests that it is difficult to achieve full coherence, at least in the context of the OECD community. For instance, Poland and Korea are the most coherent countries, with an average index of $h < 0.1$ across the 22 development modes. When looking at their highest indices, these are in the order of 0.11. Furthermore, there are no indices in the table that are positive and unambiguous at the 99\% confidence level. Perhaps this apparent boundary to the coherence index can be overcome through better specifications of the theoretical model underlying PPI or by improved estimations of the spillover networks. Nevertheless, this is a first significant improvement over existing attempts to measure policy coherence. In addition, we employ alternative distance metrics and find that our results are robust (see appendix \ref{app:robustness})

\begin{figure}[h!]
    \centering
    \caption{Coherence across development modes}
    \includegraphics[scale=.45]{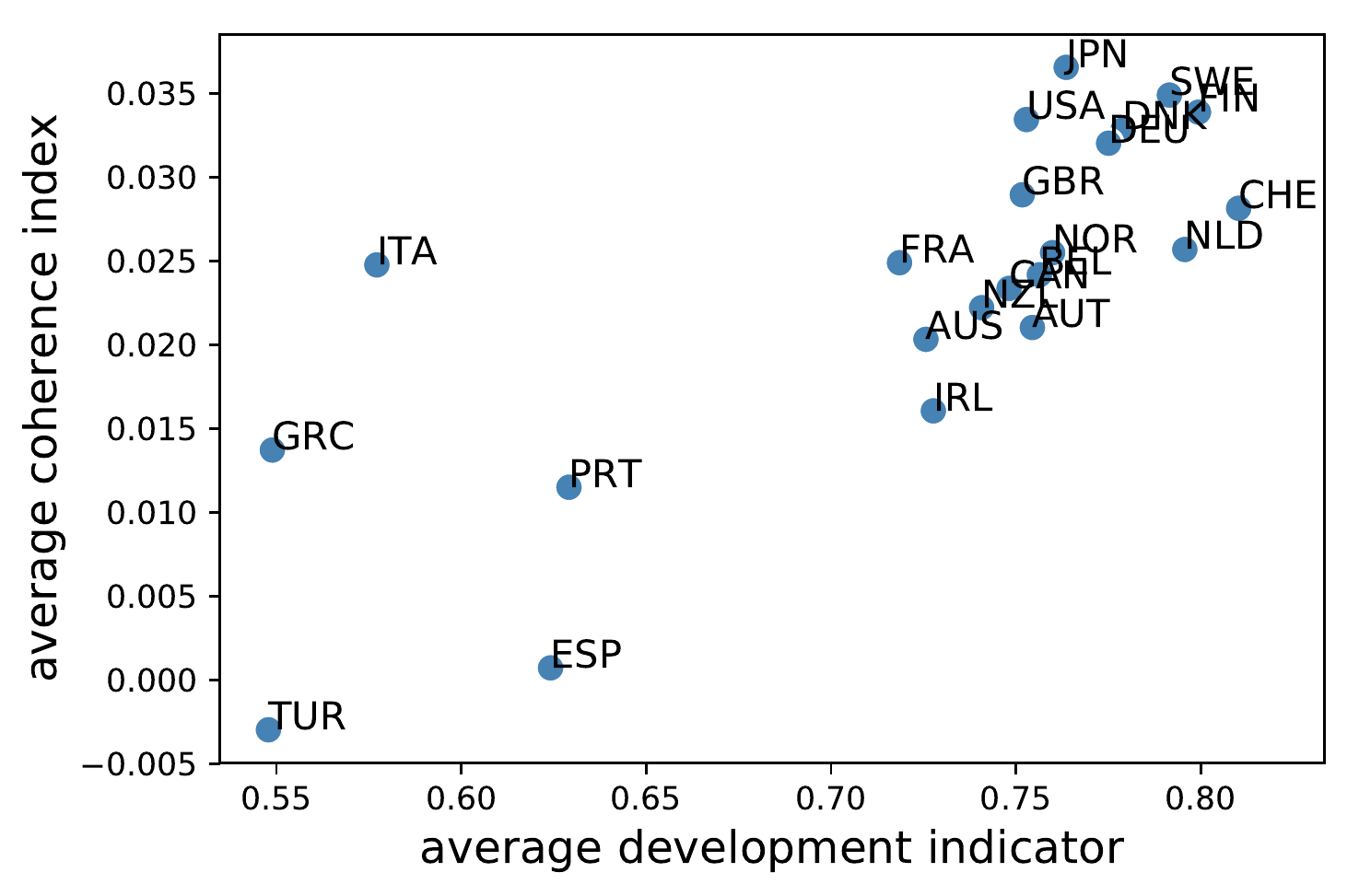}
    \includegraphics[scale=.45]{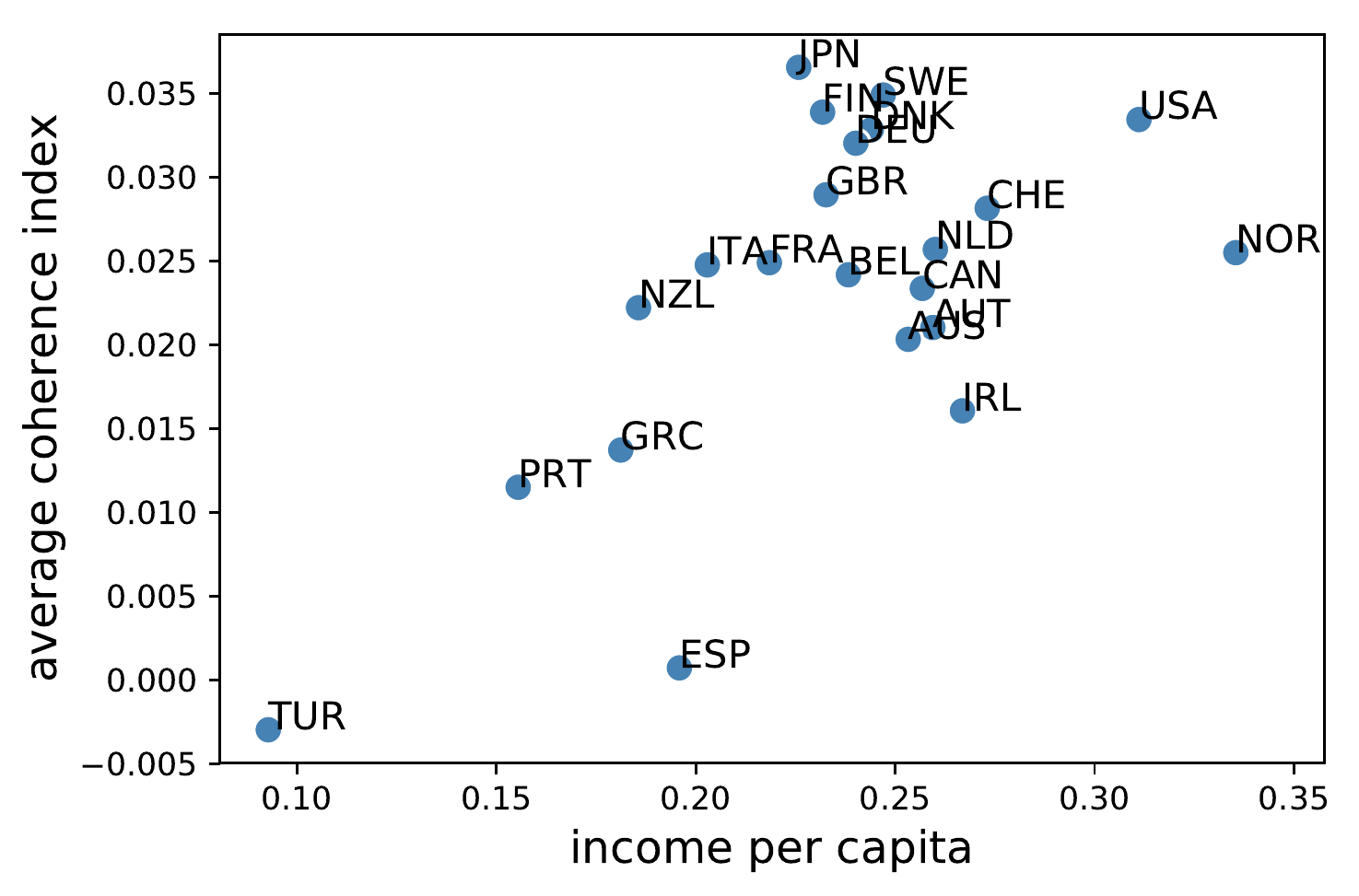}
\caption*{\footnotesize{The average is measured with the coherence index of all OECD latecomers when tracking a specific mode.}}
    \centering\label{fig:modes}
\end{figure}

Finally, using the entire sample, we would like to learn something about the average coherence of different development modes. That is, are there exemplary economies that developing countries coherently imitate in a systematic way? The left panel in Figure \ref{fig:modes} shows that this is the case. Here, we have plotted the average coherence index of each development mode against its level of performance (measured through their development indicators). The first thing to notice is that Japan, Sweden, Finland, Germany and the US are the 5 development modes that are most coherently followed. This, however, with the reservation that the overall level of the average index is still considerably low. Nevertheless, we can also observe a clear positive association between how coherently followed is a development mode and its performance. This suggests that our sample of developing countries tend to establish targets that resemble the indicators of the most advanced nations. The right panel confirms this result using income per capita as an alternative measure of performance.

\section{Discussion and conclusions}\label{sec:conclusions}

In recent years, multilateral organizations such as the UNDP, the OECD and the WB group, have actively promoted the agenda of Sustainable Development Goals (SDGs). This agenda states that countries have to implement coherent policies if they want to successfully reach a complex set of goals for the year 2030. Such complexity arises, on the one hand, from potential conflicts that emerge when multiple goals are simultaneously pursued (\textit{e.g.}, improving physical infrastructure usually damages the environment). On the other hand, complexity originates from the intricacies associated to the interactions between policies (\textit{i.e.}, from the network structure of these interactions). Clearly, meeting the 2030 agenda demands quantitative tools that can deal with such difficulties. Yet, as of today, there are no comprehensive methods to guide policymakers in this endeavour. Some commendable early attempts include building networks of development indicators and systems dynamics models. Unfortunately, these approaches present at least two of three critical shortcomings: ($i$) the requirement to gather large groups of specialized experts in many different fields, impairing scalability and replicability; ($ii$) weak validity in terms of reproducing observed statistical regularities; and ($iii$) the lack of theoretical foundations to specify detailed social mechanisms (\emph{i.e.}, the causal channels that connect policies with goals), preventing sensitivity analysis for internal validation.

In this paper, we argue that identifying the synergies and trade-offs of development indicators through network-centrality metrics is insufficient to measure policy coherence. This is so because political economy considerations are essential to disentangle the, extensively documented, technical and allocative inefficiencies. The former relates to the diversion of public funds, while the latter has to do with expenditure that is not conducive to attaining the desired development targets. In order to overcome these limitations, we propose a definition of policy coherence that takes into account the context-specific constrains imposed by the political economy. Since accounting for these constrains requires modeling specific socioeconomic processes, we propose using the Policy Priority Inference framework, built on a computational model. This allows simulating the counterfactual policy priorities that countries would establish if they were serious about attaining specific development goals. Together with a retrospective estimation, these counterfactuals are the basis for an index that quantifies the level of coherence or incoherence. Furthermore, the index can be tested for statistical significance and can be used for cross-national comparisons.   



A current limitation of our approach is that policy recommendations cannot provide point estimates on how much to prioritize each policy issue. That is, the advice is rather ordinal and identifies areas that demand interventions. Accordingly, the adoption of our proposal should be complemented with other alternatives like qualitative guidelines elaborated by field experts; especially if it is considered for the design and implementation of real-life policies. Thus, rather than trying to be overly critical with existing approaches, we would like to present our approach as an opportunity for the cross-fertilization of ideas.

Our empirical application focuses on Mexico, a developing country that joined the OECD in 1994 with the aim of, presumably, pursuing the goals and policies of the early members of this organization (as signaled by its government throughout the last 30 years). Our first finding is that Mexico's estimated policy priorities (resource allocations) for the 2006--2016 period are very different to the ones estimated in consistent profiles. Hence, its government has not been coherent despite indicating the opposite in its official discourse. Second, the Mexican index, estimated across 22 OECD development modes, does not show a positive association with the development of these nations; something that would be expected in a development strategy with a systematic emphasis on catching up with the advanced economies. Third, when comparing Mexico's retrospective and consistent allocations at the level of each policy issue (79 indicators), we find important allocative inefficiencies. In particular, there is systematic under-spending in pillars like \textit{public governance}, \textit{R\&D innovation}, \textit{education}, \textit{health}, and \textit{costs of doing business} (\textit{e.g.}, security). Yet, in topics like \text{government debt} and \textit{labor market transaction costs}, over-expenditure prevails. 

We complement the empirical analysis with two country-cases that validate our coherence index. The first case is South Korea, a very successful country that joined the OECD in 1996, and whose coherence index is positive and statistically significant in most development modes. Consistent with the literature on the Korean development history, Japan shows a high index. The second validation case is Estonia, an OECD latecomer. As expected from being a new member country, Estonia exhibits ambiguous levels of coherence. However, when looking at between-development-mode variation, Estonia shows less ambiguity and more coherence towards the Nordic countries, in particular to Denmark Finland and Sweden. This is also consistent with the literature on the role of the Nordic model in Estonian development, especially with its close ties to Finland. Japan also occupies an important place in the Estonian strategy, which is consistent with the well known technologically-oriented transformation that the Baltic nation is undergoing. 

As an additional validation test, we find a positive association between the average coherence of development modes (computed across the sample of developing countries) and their performance (income per capita or average indicators). This suggests that exemplary countries are more closely followed, confirming Akamatsu's seminal observation of the flying geese. Likewise, appendix \ref{app:triviality} demonstrates that policy coherence is not equivalent to similarity (Pearson correlation) between development indicators. This non-triviality result implies that plain benchmarking approaches, such as comparing indicators, can lead to ill-informed advice.

\bibliography{bib}

\newpage

\appendix

\section{Policy priority inference}\label{app:ppi}

Here we provide a brief explanation of PPI without extensively elaborating on the model assumptions and their justifications.

\subsection{Evolution of indicators}

There are $N$ policy issues, each with an indicator that measures its level of development. Each issue $i$ receives $P_i \in [0, 1]$ resources from the central authority, but only $C_i \in [0, Pi]$ is actually used on transformative public policies. Hence, $C_i$ measures the level of implementation efficiency in issue $i$. In this context, we interpret any inefficiency as corruption through the diversion of public funds. Hence, the gap $P_i - C_i$ is the amount of resources diverted by the agent in charge of implementing policies for issue $i$, and we say that $C_i$ is his or her contribution. In addition to the contribution of the functionary, the level of $i$'s indicator also depends on the contributions of other officials through spillover effects. We model these interdependencies as a network with adjacency matrix $A$, where $A_{ij} > 0$ if there are spillover effects from $i$ to $j$, and $A_{ij} = 0$ otherwise. As the government invests in a policy issue, its indicator grows, \textit{i.e.} the investment accumulates. This means that, if the government has set a target $T_i$ for policy issue $i$, indicator $I_i$ will reach $T_i$ after $n$ investment periods. Hence, the dynamics $I_i$ are described by

\begin{equation}
        I_{i,t} = I_{i,t-1} + \gamma(T_i-I_{i,t-1})\left(  C_{i,t} + \sum_j C_{j,t} \beta_{ji}\mathbb{A}_{ij} \right),\label{eq:propagation}
\end{equation}
where $\gamma$ is a parameter that captures the overall quality of the implemented policies in a given country (more quality implies bigger steps towards $T$).

Note that $t$ does not represent time as such, but rather a period of interactions. Hence, the number of periods it takes to converge to a target represents a number of events (\textit{e.g.}, diversions of public funds, punishments to corruption, budgetary readjustments, etc.), denoting how difficult it is for a country to reach its goals.

\subsection{Public servants}

Each public servant contributes $C_{i,t}$ to the implementation of a public policy in period $t$. How much will this official contribute depends on how costly it is to divert resources. This is determined by the benefit function

\begin{equation}
    F_{i,t} = (I_{i,t} + P_{i,t} - C_{i,t})(1 - \theta_{i,t}f_{R,t}),\label{eq:benefits}
\end{equation}
where the level of indicator $i$ gives the official political status, $\theta_{i,t}$ is an indicator function derived from the supervision of the central authority, and $f_{R,t}$ is a function mapping the indicator corresponding to the \textit{rule of law} to a probability.

The government cannot measure the real contribution of its public servants, so $P_i - C_i$ is not directly observable. However, society generates signals that the central authority may pick up in order to increase supervision efforts in specific issues. We assume that the strength of these signals is proportional to the amount of diverted public funds. Hence, we model supervision as a random variable $\theta_{i,t}$ where the outcome is 1 if the public servant in policy issue $i$ is caught diverting public funds, and zero otherwise. Then, the probability mass function of $\theta_i$ in period $t$ is

\begin{equation}
    \theta_{i,t} = 
    \begin{cases}
        1 & \text{ with probability } f_{C,t}\frac{(P_{i,t}-C_{i,t})}{\sum_{j=1}^N (P_{j,t}-C_{j,t})}, \\
        0 & \text{ with probability } 1-f_{C,t}\frac{(P_{i,t}-C_{i,t})}{\sum_{j=1}^N (P_{j,t}-C_{j,t})},
    \end{cases}\label{eq:indicator}
\end{equation}
where $f_{C,t}$ is a function mapping the indicator corresponding to the control of corruption to a probability. The indicator functions $f_{R,t}$ and $f_{C,t}$ take the form

\begin{equation}
    f_{X,t} = \frac{I_{X,t}}{e^{1-I_{X,t}}},
\end{equation}
where $X=R$ for \textit{rule of law} or $X=C$ for \textit{control of corruption}, and $I_{X,t}$ measures the level of the indicator associated to each issue.

The public servants update their contributions according to 

\begin{equation}
    C_{i,t} = \min \left\{ P_{i,t},
     \max\left(0, C_{i,t-1} +  d_{i,t} |\Delta F_{i,t}| \frac{C_{i,t-1} + C_{i,t-2}}{2} \right) \right\} \label{eq:contribution},
\end{equation}
where

\begin{equation}
    \begin{split}
        \Delta F_{i,t} = F_{i,t-1} - F_{i,t-2}\\
        \Delta C_{i,t} = C_{i,t-1} - C_{i,t-2},
    \end{split}
\end{equation}
and we define the direction of the change in benefits

\begin{equation}
    d_{i,t} = \text{sgn} (\Delta F_{i,t} \cdot \Delta C_{i,t}).
\end{equation}

For consistency, the \textit{min} and \textit{max} functions bound the public servants' contributions to the interval $[0, P_{i,t}]$.

\subsection{Central authority}

The central authority has a vector of targets $(T_1, \dots ,T_N)$ that it wants to achieve for its development indicators. The government's problem is deciding how to best allocate its limited resources to different policies, in order to reduce the gap between the current indicators and the targets. Formally, the problem is 

\begin{equation}
    \min \left[ \sum^N_{i=1} \left( I_{i,t}-T_{i} \right)^2  \right]^{\frac{1}{2}}\label{eq:MSE}
\end{equation}

Equation \ref{eq:indicator} indicates that $I_{i,t}$ is a function of the resource allocation; therefore, $P_{1,t}, \dots , P_{N,t}$ are the control variables of the central authority. We call a specific configuration of these variables an \textit{allocation profile}. In addition, the amount of resources that the government can invest is restricted by

\begin{equation}
    \sum_i^N P_{i,t} \leq B \; \forall \; t.\label{eq:budget}
\end{equation}
where $B$ reflects the amount of non-committed resources of the central authority (this excludes current expenditure). Thus, empirically speaking, $B$ must be chosen such that it reflects how much budget countries can spare in transforming their economies through public policy.

Each period, the central authority determines an allocation profile and evaluates the gap between the targets and indicators. The amount of resources allocated to policy issue $i$ is determined by 

\begin{equation}
    p_{i,t} = \frac{q_{i,t}}{\sum_j^N q_{i,t}},
\label{eq:allocation}
\end{equation}
where $q_{i,t}$ is the propensity to assign resources to policy $i$, defined as

\begin{equation}
    q_{i,t} = (T_i-I_{i,t})(K_i+1)(1-\theta_{i,t}f_{R,t}),\label{eq:propensity}
\end{equation}
where $K_i$ is the number of connections of node $i$, also known as its degree. 

Equation \ref{eq:propensity} summarizes the intuition of how governments learn and adapt their policy priorities.  First, the government tries to close the gap $T_i-I_{i,t}$ between the target and the indicator in order to solve equation \ref{eq:MSE}. Second, $K_i$ is a proxy about how critical a policy issue is. That is, policy issues with a large $K_i$ are central to the development process because of their high inter-connectivity to other issues. Third, the government tries to reach its targets while, at the same time, attempting to discourage corruption through budgetary readjustments.

Finally, the amount of resources allocated to policy issue $i$ is 

\begin{equation}
    P_{i,t} = p_{i,t}B.
\end{equation}

In summary, the model generates endogenous indicators from a political economy game in which policy issues are interdependent. The misalignment between the incentives of the central authority and those of the public servants elicits free-riding and illicit personal gains. In order to reach its goals, the government penalizes corruption and assigns resources to policy issues with more potential for improving overall economic performance. Its data sources are the initial level of a country's indicator, some desired targets, a spillover network and a budget constrain. Parameter $\gamma$ is estimated by fitting the model to the observed levels of corruption (see section \ref{sec:methods}). By simulating the evolution of development indicators, we infer policy priorities through the allocation profile expected from Monte Carlo simulations.

\section{Model calibration}\label{app:calibration}

In order to estimate $\gamma$ from equation \ref{eq:propagation}, we fit the model's corruption output to an independent indicator of diversion of public funds. This strategy seeks to exploit the cross-national variation of corruption, so it is necessary to run the model for all countries. Thus, we calibrate the model for each country using the development indicators of 2006 as initial conditions; the indicators of 2016 as the targets; an indicator of public expenditure as a fraction of GDP as the budget constraint; and the estimated spillover networks.

The output variable of interest to the calibration is the total amount of corruption

\begin{equation}
    \bar{D} = \frac{1}{NB}\sum_{i=1}^N \sum^\ell_{t=0} (P_{i,t} - C_{i,t})\label{eq:data_corruption}
\end{equation}
where $l$ is the number of periods it took for a simulation to converge. Recall that a period in the model does not represent time. In fact, two countries that are calibrated for the same sample period may have very different $l$s. Since $l$ represents the frequency of events before convergence, it also reflects the incidence of corruption. Therefore, $\bar{D}$ measures the expected level of corruption across policy issues in a given country. For a single country, we compute the expected $D$ across a Monte Carlo sample. Doing this for all countries gives us the cross-national relative difference in diversion of public funds. The estimation procedure minimizes the distance between this quantity and the empirical one. It finds a set of parameters $\gamma$ to classify all countries while controlling for overfitting. This is a standard method in classification problems, commonly known as finding the true number of clusters in a dataset, and its closest analogy in linear regression would be the problem of parameter heterogeneity across sub-samples (see \cite{castaneda_how_2018} for further details).

\begin{figure}[h!]
    \centering
    \caption{Model calibration}
    \includegraphics[scale=.45]{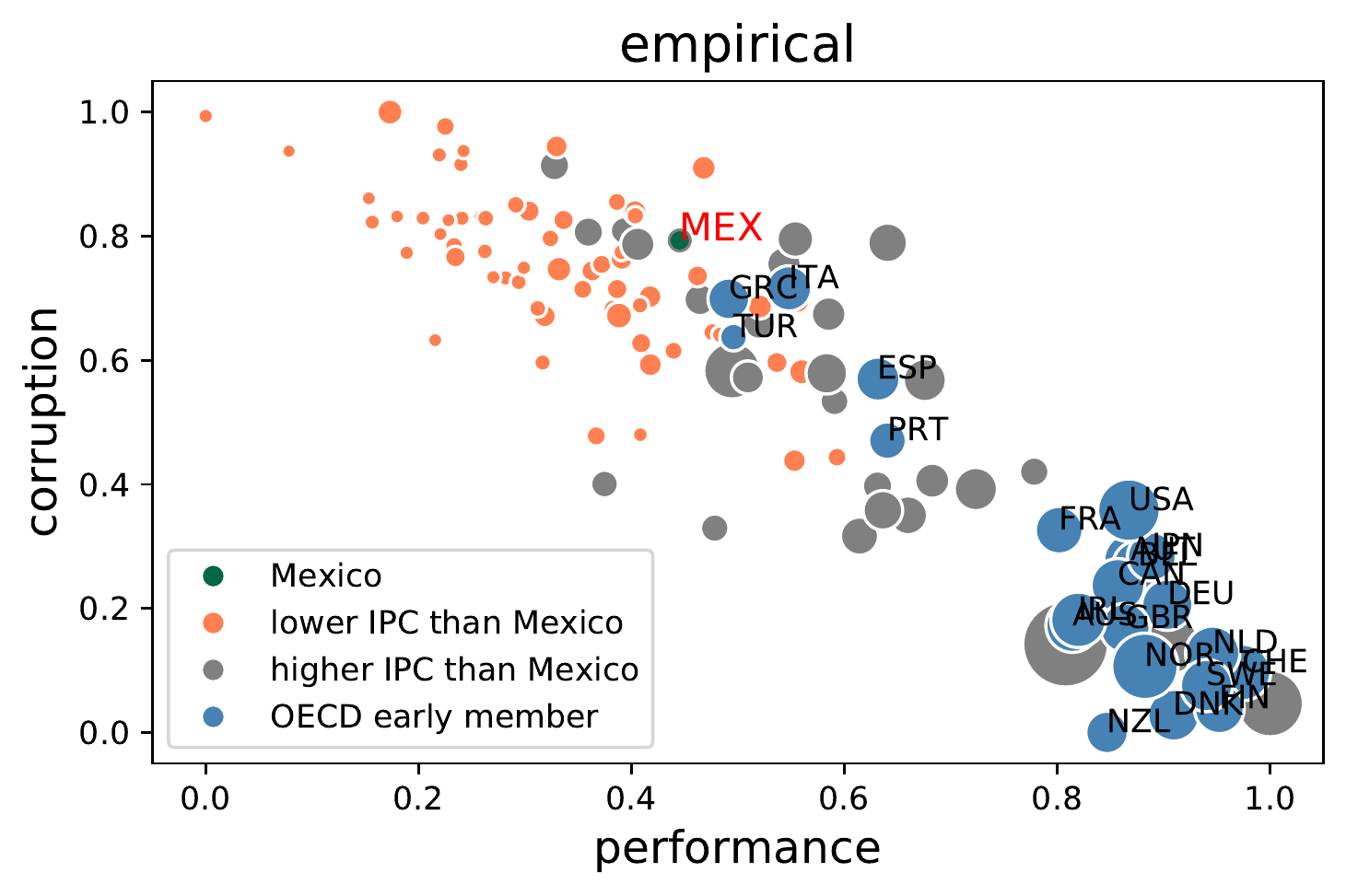}
    \includegraphics[scale=.45]{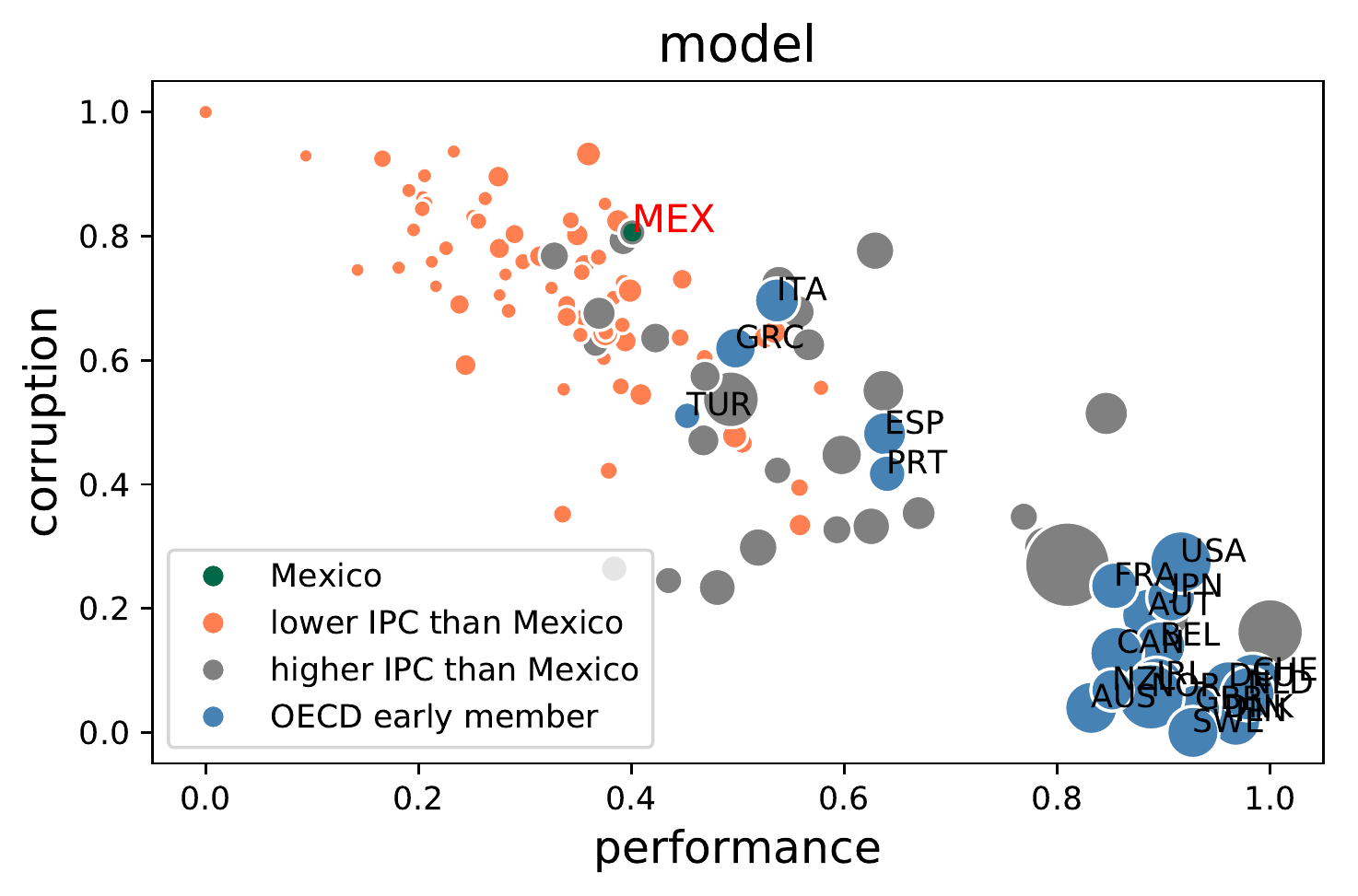}
\caption*{\footnotesize{Each dot in panel (a) corresponds to an 11-years average of a country. The Y-axis corresponds to the average for diversion of public funds, directly obtained from an indicator in the data set under the development pillar of public governance. The X-axis is the arithmetic mean of the rest of the indicators. The coordinates for the dots in panel (b) are computed from the model’s outputs from 1000 Monte Carlo simulations for each country. Marker sizes denote income per capita (logarithmic).}}
    \centering\label{fig:calibration}
\end{figure}

Figure 3 shows the empirical and estimated levels of corruption. As an informal validation, we plot the well known relationship between corruption and the level of development of a country (see \cite{castaneda_how_2018} for multiple formal validation tests). In this case, the level of development is measured through the average level of a country's indicators. Note that the model has not been fitted for performance, so reorderings on the horizontal axis are expected.

\section{Non-triviality of the coherence index}\label{app:triviality}

\begin{figure}[h!]
    \centering
    \caption{The irrelevance of development-indicator comparisons for measuring policy coherence}
    \includegraphics[scale=.45]{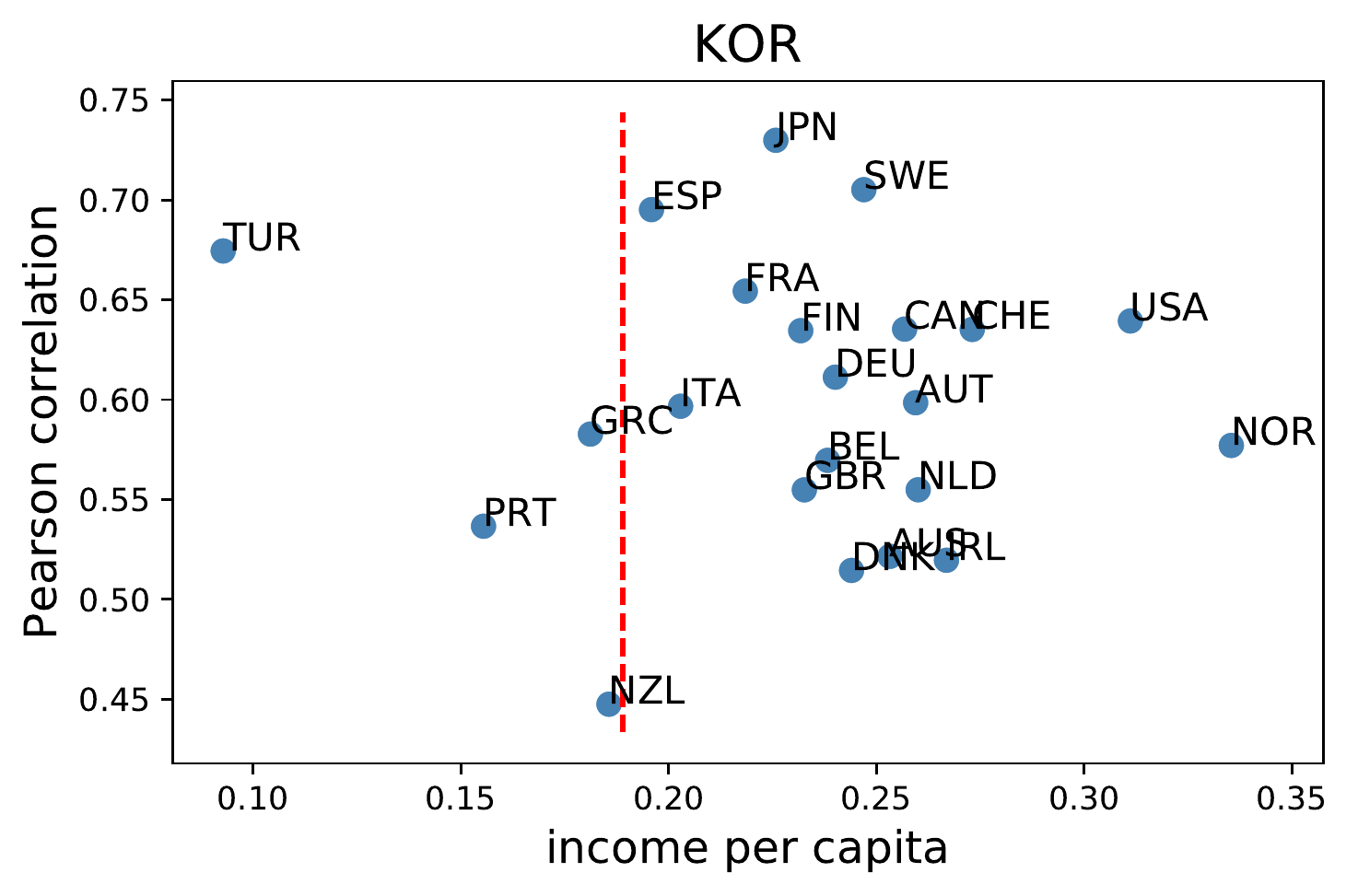}
    \includegraphics[scale=.45]{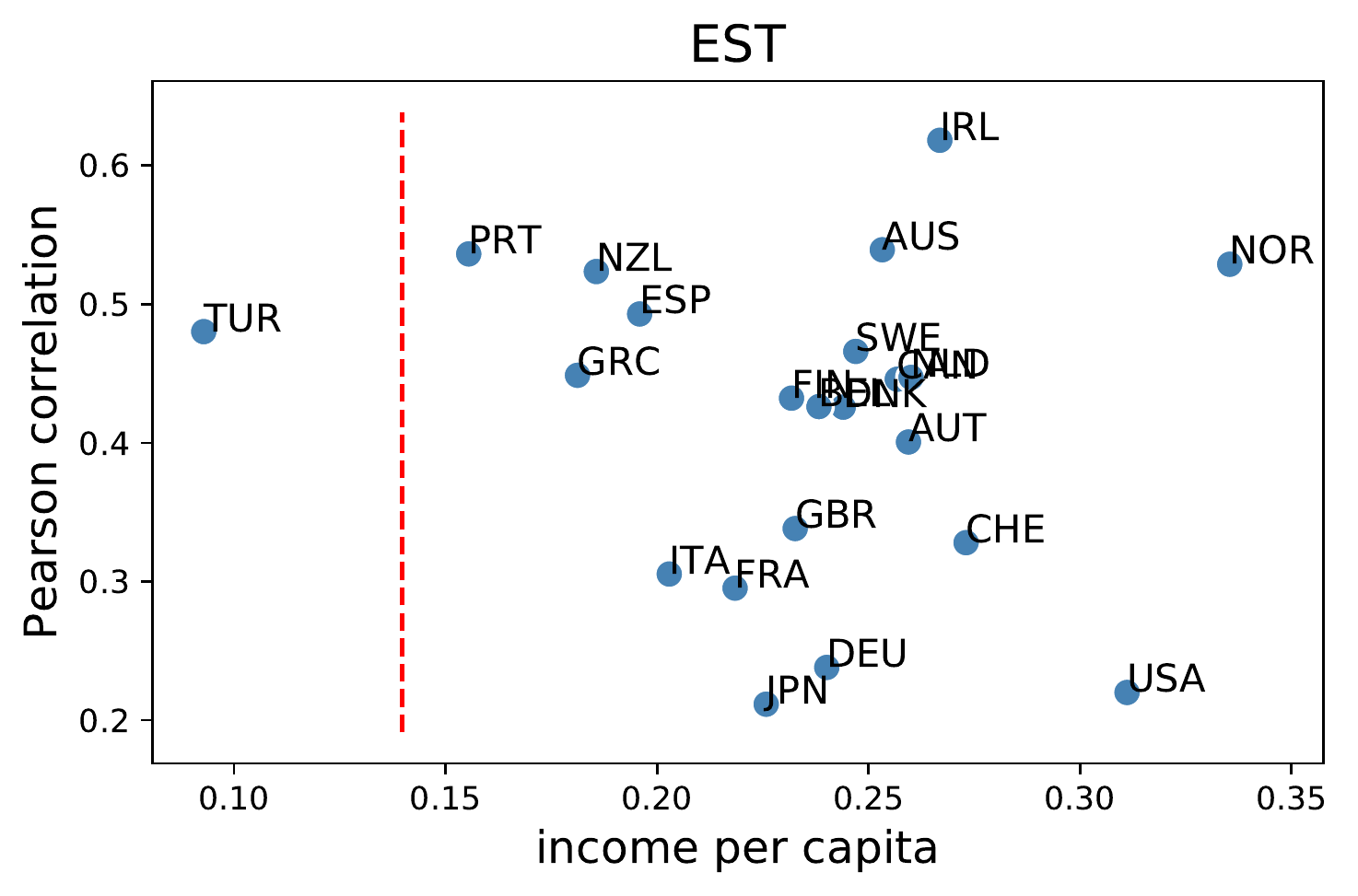}
    \includegraphics[scale=.45]{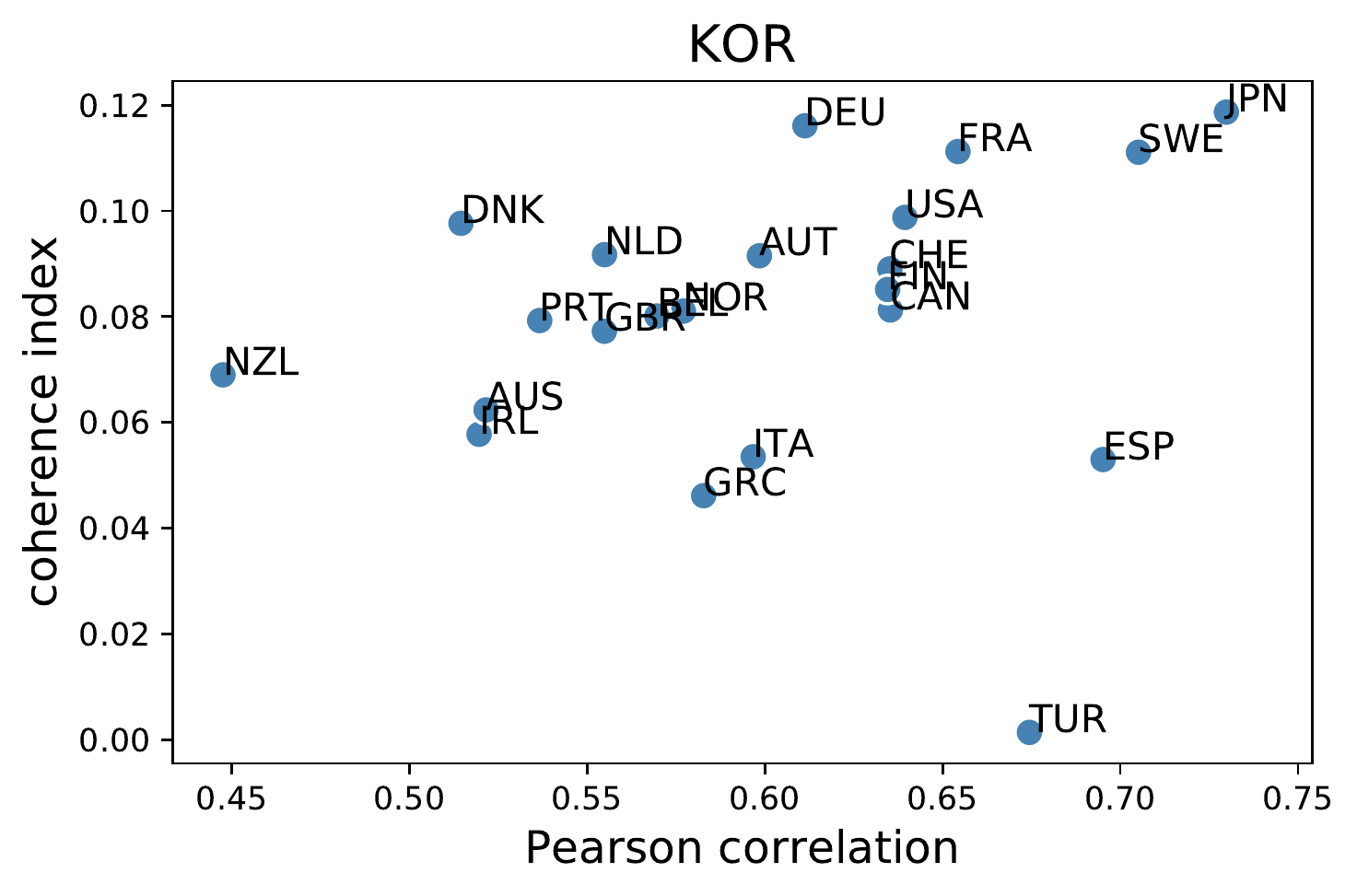}
    \includegraphics[scale=.45]{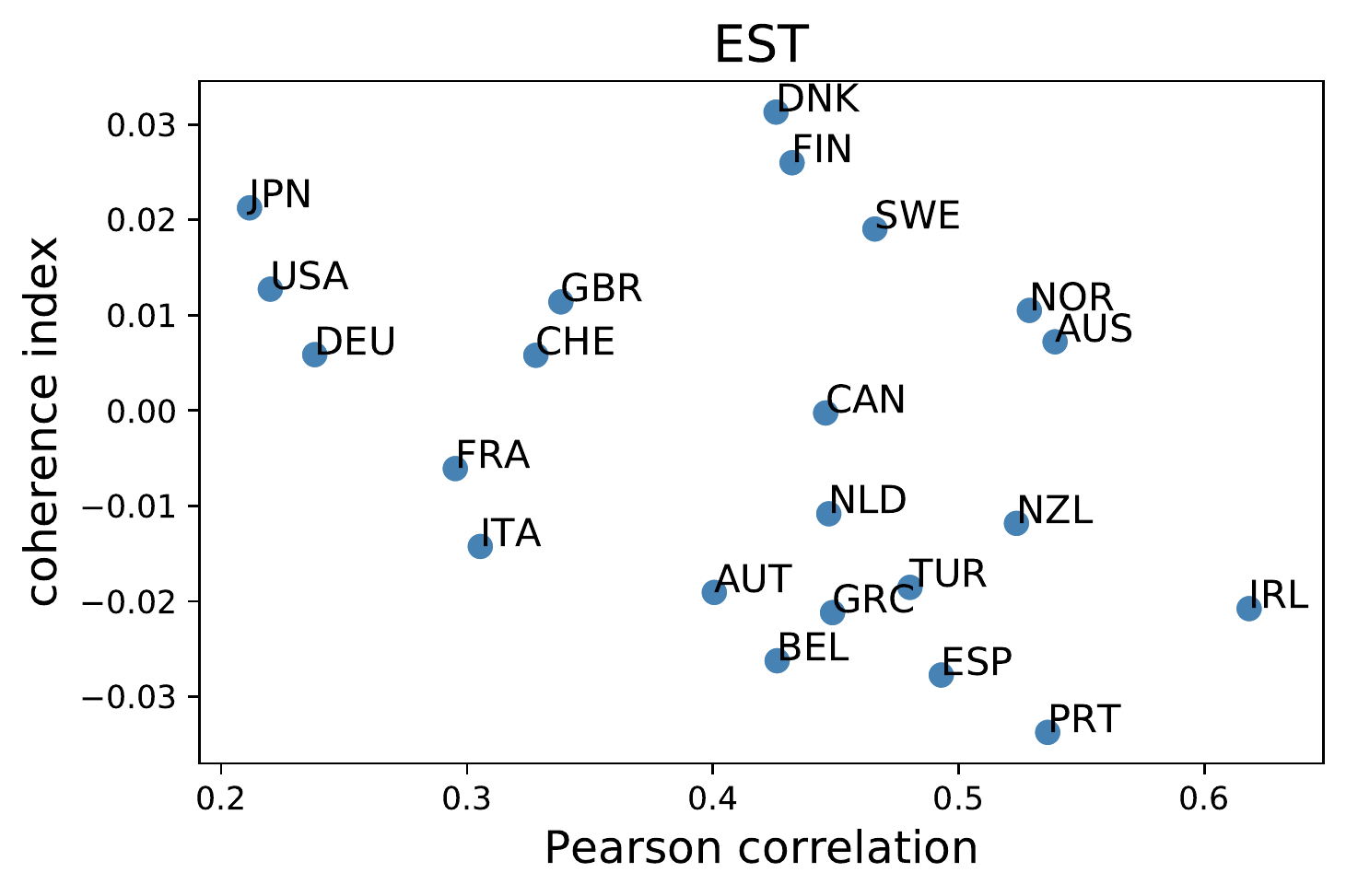}

\caption*{\footnotesize{The dashed red line denotes the income per capita of the developing country under analysis. We use the Pearson correlation as a similarity index between the country's development indicators and those of its potential development modes}}
    \centering\label{fig:non-triviality}
\end{figure}

In this appendix, we show that policy coherence is not equivalent to similarity between development indicators. This only happen when the two sets of indicators are almost identical in terms of order and magnitude. The top panels in Figure \ref{fig:non-triviality} try to replicate the results from our validation cases in Figure \ref{fig:validation}, using the Pearson correlation coefficient as a measure for similarity between the retrospective and the counterfactual targets (the results are robust for other similarity metrics). If similarity between targets explains coherence, then the top panels should look identical to the ones in Figure \ref{fig:validation}. Clearly this is not the case; for example, the German indicators are considerably less similar to the Korean ones than the Japanese indicators and, yet, the European country is the second most coherent mode for Korea. On the other hand Ireland is the the country with most similar indicators to Estonia, but its corresponding index falls into the domain of incoherence.

For the Korean case, however, Japan is both the most similar and the most coherent. Naturally, this raises the question of whether, among high levels of similarity between indicators, it is possible to find a positive association with coherence. The bottom panels in Figure \ref{fig:non-triviality} show that this is not the case. For instance, Korean development modes like Spain and Turkey with a high degree of similarity present also a low coherence index with respecto to the Asian country. Then, in the Estonian case, there are 9 development modes with a similarity above 0.4, but whose coherence index with the Baltic country is negative. In brief, the bottom panels do not show a positive relationship between the similarity and the coherence indices.

This strengthens our advocacy for computational models that capture the policymaking process in order to infer policy priorities. It also shows the virtues of PPI, and points out to the flaws that naive benchmarking approaches such as comparing indicators can have. Thus, our work highlights the need to combine data-driven tools with theoretically-founded analyses to deal with the problem of measuring the coherence of policy priorities.

\section{Robustness}\label{app:robustness}

Tables \ref{tab:cosine}, \ref{tab:correlation} and \ref{tab:euclidean} present estimates of the coherence index for the augmented sample under three alternative distance metrics: cosine, correlation and Euclidean distances. Overall, these tables demonstrate the robustness of our results. For example, Mexico us persistently not coherent; Japan occupies a prominent position among Korea's most-coherent development modes; and Estonia is most-coherent with the Nordic countries. These three alternative (and popular) distance metrics are defined as follows.

Cosine distance is defined as

\begin{equation}
    1 - \frac{X \cdot Y}{||X||_2||Y||_2},
\end{equation}
where $||\cdot||_2$ is the L2 norm $\left(\sum x^2 \right)^{1/2}$. The correlation distance is

\begin{equation}
    1 - \frac{(X-\bar{X}) \cdot (Y-\bar{Y})}{||(X-\bar{X})||_2 ||(Y-\bar{Y})||_2},
\end{equation}
and the Euclidean distance is the L2 norm for the difference between vectors $X$ and $Y$:

\begin{equation}
    \left(\sum_i (X_i - Y_i)^2 \right)^{1/2}.
\end{equation}

\begin{table}[h!]
    \caption{Coherence indices using cosine distance}
    \footnotesize
    \begin{center}
        \begin{tabular}{ c c c c c c c c c c c c c c}
            \hline
            Mode & CHL & CZE & EST & HUN & ISR & KOR & LTU & LVA & MEX & POL & SVK & SVN\\
            \hline
AUS&$0.17^{*}$&$0.01^{}$&$0.0^{}$&$-0.04^{}$&$-0.04^{}$&$0.1^{}$&$0.08^{}$&$-0.04^{}$&$-0.03^{}$&$0.15^{*}$&$-0.08^{**}$&$0.05^{}$\\
AUT&$0.15^{}$&$0.09^{}$&$-0.02^{}$&$-0.06^{}$&$-0.05^{}$&$0.15^{*}$&$0.0^{}$&$-0.01^{}$&$-0.05^{}$&$0.15^{*}$&$-0.06^{}$&$0.07^{}$\\
BEL&$0.08^{}$&$0.16^{*}$&$-0.04^{}$&$0.02^{}$&$-0.04^{}$&$0.13^{}$&$0.04^{}$&$-0.03^{}$&$-0.04^{}$&$0.16^{*}$&$-0.04^{}$&$0.06^{}$\\
CAN&$0.15^{}$&$0.03^{}$&$-0.01^{}$&$-0.04^{}$&$-0.05^{}$&$0.13^{}$&$0.03^{}$&$-0.02^{}$&$-0.04^{}$&$0.15^{}$&$-0.05^{}$&$0.07^{}$\\
CHE&$0.13^{}$&$0.09^{}$&$-0.0^{}$&$-0.03^{}$&$-0.03^{}$&$0.13^{}$&$0.05^{}$&$-0.02^{}$&$-0.05^{}$&$0.11^{}$&$-0.05^{}$&$0.08^{}$\\
DEU&$0.15^{}$&$0.15^{*}$&$0.0^{}$&$-0.01^{}$&$-0.04^{}$&$0.16^{*}$&$0.01^{}$&$-0.02^{}$&$-0.05^{}$&\cellcolor{blue!25}$0.17^{*}$&$-0.07^{**}$&$0.07^{}$\\
DNK&$0.14^{}$&$0.12^{}$&$0.04^{}$&$-0.05^{}$&$-0.03^{}$&$0.16^{*}$&$0.08^{}$&$-0.02^{}$&$-0.03^{}$&$0.13^{}$&$-0.06^{}$&$0.08^{}$\\
ESP&$0.05^{}$&$-0.01^{}$&$-0.04^{}$&$-0.08^{}$&$-0.05^{*}$&$0.1^{}$&$0.02^{}$&$-0.04^{}$&$-0.05^{}$&$0.15^{*}$&$-0.04^{}$&$0.01^{}$\\
FIN&$0.15^{}$&$0.14^{*}$&\cellcolor{blue!25}$0.05^{}$&$-0.0^{}$&$-0.03^{}$&$0.12^{}$&$0.06^{}$&$-0.03^{}$&$-0.04^{}$&$0.14^{}$&$-0.07^{*}$&$0.07^{}$\\
FRA&$0.1^{}$&\cellcolor{blue!25}$0.16^{*}$&$-0.01^{}$&$-0.0^{}$&$-0.03^{}$&\cellcolor{blue!25}$0.19^{*}$&$0.0^{}$&$-0.05^{}$&$-0.04^{}$&$0.16^{*}$&$-0.05^{*}$&$0.06^{}$\\
GBR&$0.17^{*}$&$0.11^{}$&$0.01^{}$&$-0.05^{}$&$-0.03^{}$&$0.11^{}$&$0.03^{}$&$-0.05^{}$&$-0.03^{}$&$0.17^{*}$&$-0.06^{*}$&$0.07^{}$\\
GRC&$0.0^{}$&$-0.0^{}$&$-0.02^{}$&\cellcolor{blue!25}$0.14^{}$&$-0.02^{}$&$0.11^{}$&$-0.05^{}$&$0.05^{}$&$-0.02^{}$&$0.08^{}$&$0.06^{}$&$0.01^{}$\\
IRL&$0.13^{}$&$0.07^{}$&$-0.03^{}$&$-0.03^{}$&$-0.03^{}$&$0.11^{}$&$0.01^{}$&$-0.04^{}$&$-0.04^{}$&$0.15^{*}$&$-0.06^{**}$&$0.04^{}$\\
ITA&$0.03^{}$&$0.0^{}$&$-0.03^{}$&$0.05^{}$&$-0.01^{}$&$0.12^{}$&\cellcolor{blue!25}$0.13^{}$&\cellcolor{blue!25}$0.09^{}$&\cellcolor{blue!25}$-0.0^{}$&$0.08^{}$&\cellcolor{blue!25}$0.1^{}$&$0.04^{}$\\
JPN&$0.1^{}$&$0.12^{}$&$0.04^{}$&$0.02^{}$&$-0.02^{}$&$0.18^{*}$&$0.1^{}$&$0.03^{}$&$-0.05^{}$&$0.13^{}$&$-0.02^{}$&$0.07^{}$\\
NLD&$0.15^{}$&$0.07^{}$&$-0.02^{}$&$-0.04^{}$&$-0.03^{}$&$0.15^{*}$&$0.04^{}$&$-0.02^{}$&$-0.04^{}$&$0.16^{*}$&$-0.06^{}$&$0.07^{}$\\
NOR&\cellcolor{blue!25}$0.2^{*}$&$0.05^{}$&$0.01^{}$&$-0.04^{}$&$-0.04^{}$&$0.15^{*}$&$0.06^{}$&$-0.02^{}$&$-0.05^{}$&$0.13^{}$&$-0.06^{*}$&$0.07^{}$\\
NZL&$0.16^{}$&$0.02^{}$&$-0.02^{}$&$-0.02^{}$&$-0.02^{}$&$0.13^{*}$&$0.08^{}$&$-0.03^{}$&$-0.02^{}$&$0.15^{*}$&$-0.06^{}$&$0.05^{}$\\
PRT&$0.04^{}$&$0.03^{}$&$-0.05^{*}$&$0.01^{}$&$-0.01^{}$&$0.17^{*}$&$0.0^{}$&$-0.01^{}$&$-0.05^{**}$&$0.11^{}$&$-0.02^{}$&$0.03^{}$\\
SWE&$0.19^{*}$&$0.12^{}$&$0.03^{}$&$-0.04^{}$&$-0.03^{}$&$0.19^{*}$&$0.06^{}$&$-0.03^{}$&$-0.05^{}$&$0.16^{*}$&$-0.06^{*}$&\cellcolor{blue!25}$0.08^{}$\\
TUR&$0.02^{}$&$-0.03^{*}$&$-0.03^{}$&$-0.04^{}$&$-0.02^{}$&$0.01^{}$&$0.03^{}$&$-0.07^{}$&$-0.02^{}$&$0.09^{}$&$0.02^{}$&$0.02^{}$\\
USA&$0.16^{*}$&$0.13^{}$&$0.02^{}$&$-0.05^{}$&\cellcolor{blue!25}$-0.01^{}$&$0.14^{*}$&$0.09^{}$&$-0.03^{}$&$-0.03^{}$&$0.16^{*}$&$-0.04^{}$&$0.08^{}$\\
\hline
Average&$0.12$&$0.07$&$-0.01$&$-0.02$&$-0.03$&$0.13$&$0.04$&$-0.02$&$-0.04$&$0.14$&$-0.04$&$0.0$ \\
            \hline
        \end{tabular}
    \end{center}\label{tab:cosine}
    \caption*{\footnotesize{Shaded areas are the development modes with maximum coherence for each country. Stars indicate non-ambiguity significance. *:$p<0.1$, **:$p<0.05$}}
\end{table}

\begin{table}[h!]
    \caption{Coherence indices using correlation distance}
    \footnotesize
    \begin{center}
        \begin{tabular}{ c c c c c c c c c c c c c c}
            \hline
            Mode & CHL & CZE & EST & HUN & ISR & KOR & LTU & LVA & MEX & POL & SVK & SVN\\
            \hline
AUS&$0.18^{*}$&$0.01^{}$&$0.0^{}$&$-0.04^{}$&$-0.04^{}$&$0.1^{}$&$0.08^{}$&$-0.05^{}$&$-0.03^{}$&$0.15^{*}$&$-0.08^{**}$&$0.05^{}$\\
AUT&$0.15^{}$&$0.09^{}$&$-0.02^{}$&$-0.07^{}$&$-0.05^{}$&$0.15^{*}$&$0.0^{}$&$-0.01^{}$&$-0.05^{}$&$0.15^{*}$&$-0.07^{}$&$0.07^{}$\\
BEL&$0.08^{}$&$0.16^{*}$&$-0.04^{}$&$0.02^{}$&$-0.04^{}$&$0.13^{}$&$0.04^{}$&$-0.03^{}$&$-0.04^{}$&$0.16^{*}$&$-0.04^{}$&$0.06^{}$\\
CAN&$0.15^{}$&$0.03^{}$&$-0.01^{}$&$-0.04^{}$&$-0.05^{}$&$0.13^{}$&$0.03^{}$&$-0.02^{}$&$-0.04^{}$&$0.16^{}$&$-0.05^{}$&$0.07^{}$\\
CHE&$0.13^{}$&$0.09^{}$&$-0.0^{}$&$-0.03^{}$&$-0.03^{}$&$0.13^{}$&$0.05^{}$&$-0.02^{}$&$-0.05^{}$&$0.12^{}$&$-0.05^{}$&$0.08^{}$\\
DEU&$0.15^{}$&$0.16^{*}$&$0.0^{}$&$-0.01^{}$&$-0.04^{}$&$0.16^{*}$&$0.01^{}$&$-0.02^{}$&$-0.05^{}$&\cellcolor{blue!25}$0.18^{*}$&$-0.07^{**}$&$0.07^{}$\\
DNK&$0.14^{}$&$0.12^{}$&$0.05^{}$&$-0.05^{}$&$-0.03^{}$&$0.16^{*}$&$0.08^{}$&$-0.02^{}$&$-0.03^{}$&$0.14^{}$&$-0.07^{}$&$0.08^{}$\\
ESP&$0.05^{}$&$-0.01^{}$&$-0.04^{}$&$-0.08^{}$&$-0.05^{*}$&$0.1^{}$&$0.02^{}$&$-0.04^{}$&$-0.05^{}$&$0.15^{*}$&$-0.04^{}$&$0.01^{}$\\
FIN&$0.15^{}$&$0.15^{*}$&\cellcolor{blue!25}$0.05^{}$&$-0.0^{}$&$-0.03^{}$&$0.12^{}$&$0.07^{}$&$-0.03^{}$&$-0.05^{}$&$0.14^{}$&$-0.07^{*}$&$0.07^{}$\\
FRA&$0.1^{}$&\cellcolor{blue!25}$0.16^{*}$&$-0.01^{}$&$-0.0^{}$&$-0.03^{}$&\cellcolor{blue!25}$0.19^{*}$&$0.0^{}$&$-0.05^{}$&$-0.05^{}$&$0.16^{*}$&$-0.05^{*}$&$0.06^{}$\\
GBR&$0.17^{*}$&$0.11^{}$&$0.01^{}$&$-0.05^{}$&$-0.03^{}$&$0.12^{}$&$0.03^{}$&$-0.05^{}$&$-0.04^{}$&$0.17^{*}$&$-0.06^{*}$&$0.07^{}$\\
GRC&$0.0^{}$&$-0.0^{}$&$-0.02^{}$&\cellcolor{blue!25}$0.14^{}$&$-0.02^{}$&$0.11^{}$&$-0.05^{}$&$0.05^{}$&$-0.02^{}$&$0.08^{}$&$0.06^{}$&$0.01^{}$\\
IRL&$0.13^{}$&$0.07^{}$&$-0.03^{}$&$-0.03^{}$&$-0.04^{}$&$0.11^{}$&$0.01^{}$&$-0.05^{}$&$-0.04^{}$&$0.15^{*}$&$-0.06^{**}$&$0.04^{}$\\
ITA&$0.03^{}$&$0.0^{}$&$-0.03^{}$&$0.05^{}$&$-0.01^{}$&$0.12^{}$&\cellcolor{blue!25}$0.13^{}$&\cellcolor{blue!25}$0.1^{}$&\cellcolor{blue!25}$-0.0^{}$&$0.08^{}$&\cellcolor{blue!25}$0.1^{}$&$0.04^{}$\\
JPN&$0.1^{}$&$0.12^{}$&$0.04^{}$&$0.02^{}$&$-0.02^{}$&$0.18^{*}$&$0.1^{}$&$0.03^{}$&$-0.05^{}$&$0.13^{}$&$-0.02^{}$&$0.07^{}$\\
NLD&$0.15^{}$&$0.08^{}$&$-0.02^{}$&$-0.04^{}$&$-0.03^{}$&$0.15^{*}$&$0.04^{}$&$-0.02^{}$&$-0.04^{}$&$0.16^{*}$&$-0.06^{}$&$0.07^{}$\\
NOR&\cellcolor{blue!25}$0.2^{*}$&$0.05^{}$&$0.01^{}$&$-0.04^{}$&$-0.04^{}$&$0.15^{*}$&$0.06^{}$&$-0.02^{}$&$-0.05^{}$&$0.13^{}$&$-0.06^{*}$&$0.07^{}$\\
NZL&$0.17^{}$&$0.02^{}$&$-0.02^{}$&$-0.02^{}$&$-0.02^{}$&$0.13^{*}$&$0.09^{}$&$-0.03^{}$&$-0.02^{}$&$0.16^{*}$&$-0.06^{}$&$0.05^{}$\\
PRT&$0.04^{}$&$0.03^{}$&$-0.05^{*}$&$0.01^{}$&$-0.01^{}$&$0.17^{*}$&$0.0^{}$&$-0.01^{}$&$-0.05^{**}$&$0.12^{}$&$-0.02^{}$&$0.03^{}$\\
SWE&$0.2^{*}$&$0.12^{}$&$0.03^{}$&$-0.04^{}$&$-0.03^{}$&$0.19^{*}$&$0.06^{}$&$-0.03^{}$&$-0.05^{}$&$0.17^{*}$&$-0.07^{*}$&\cellcolor{blue!25}$0.08^{}$\\
TUR&$0.02^{}$&$-0.03^{*}$&$-0.03^{}$&$-0.04^{}$&$-0.02^{}$&$0.01^{}$&$0.03^{}$&$-0.07^{}$&$-0.02^{}$&$0.09^{}$&$0.02^{}$&$0.02^{}$\\
USA&$0.16^{*}$&$0.13^{}$&$0.02^{}$&$-0.05^{}$&\cellcolor{blue!25}$-0.01^{}$&$0.14^{*}$&$0.09^{}$&$-0.04^{}$&$-0.03^{}$&$0.16^{*}$&$-0.04^{}$&$0.08^{}$\\
\hline
Average&$0.12$&$0.07$&$-0.01$&$-0.02$&$-0.03$&$0.13$&$0.04$&$-0.02$&$-0.04$&$0.14$&$-0.04$&$0.0$ \\
            \hline
        \end{tabular}
    \end{center}\label{tab:correlation}
    \caption*{\footnotesize{Shaded areas are the development modes with maximum coherence for each country. Stars indicate non-ambiguity significance. *:$p<0.1$, **:$p<0.05$}}
\end{table}

\begin{table}[h!]
    \caption{Coherence indices using Euclidean distance}
    \footnotesize
    \begin{center}
        \begin{tabular}{ c c c c c c c c c c c c c c}
            \hline
            Mode & CHL & CZE & EST & HUN & ISR & KOR & LTU & LVA & MEX & POL & SVK & SVN\\
            \hline
AUS&$0.09^{*}$&$0.0^{}$&$0.0^{}$&$-0.02^{}$&$-0.02^{}$&$0.05^{}$&$0.03^{}$&$-0.02^{}$&$-0.01^{}$&$0.07^{*}$&$-0.03^{**}$&$0.02^{}$\\
AUT&$0.07^{}$&$0.04^{}$&$-0.01^{}$&$-0.03^{}$&$-0.03^{}$&$0.07^{*}$&$0.0^{}$&$-0.0^{}$&$-0.02^{}$&$0.07^{*}$&$-0.03^{}$&$0.03^{}$\\
BEL&$0.04^{}$&\cellcolor{blue!25}$0.08^{*}$&$-0.02^{}$&$0.01^{}$&$-0.02^{}$&$0.07^{}$&$0.02^{}$&$-0.01^{}$&$-0.01^{}$&$0.08^{*}$&$-0.02^{}$&$0.03^{}$\\
CAN&$0.07^{}$&$0.01^{}$&$-0.01^{}$&$-0.02^{}$&$-0.02^{}$&$0.06^{}$&$0.01^{}$&$-0.01^{}$&$-0.01^{}$&$0.07^{}$&$-0.02^{}$&$0.03^{}$\\
CHE&$0.06^{}$&$0.04^{}$&$-0.0^{}$&$-0.02^{}$&$-0.01^{}$&$0.06^{}$&$0.02^{}$&$-0.01^{}$&$-0.02^{}$&$0.05^{}$&$-0.02^{}$&$0.04^{}$\\
DEU&$0.07^{}$&$0.07^{*}$&$0.0^{}$&$-0.0^{}$&$-0.02^{}$&$0.08^{*}$&$0.01^{}$&$-0.01^{}$&$-0.02^{}$&\cellcolor{blue!25}$0.08^{*}$&$-0.03^{**}$&$0.03^{}$\\
DNK&$0.07^{}$&$0.06^{}$&$0.02^{}$&$-0.02^{}$&$-0.01^{}$&$0.08^{*}$&$0.04^{}$&$-0.01^{}$&$-0.01^{}$&$0.06^{}$&$-0.02^{}$&$0.04^{}$\\
ESP&$0.03^{}$&$-0.0^{}$&$-0.02^{}$&$-0.04^{}$&$-0.03^{*}$&$0.05^{}$&$0.01^{}$&$-0.02^{}$&$-0.02^{}$&$0.07^{*}$&$-0.02^{}$&$0.01^{}$\\
FIN&$0.07^{}$&$0.07^{*}$&\cellcolor{blue!25}$0.02^{}$&$-0.0^{}$&$-0.01^{}$&$0.06^{}$&$0.03^{}$&$-0.01^{}$&$-0.02^{}$&$0.07^{}$&$-0.03^{*}$&$0.04^{}$\\
FRA&$0.05^{}$&$0.07^{*}$&$-0.01^{}$&$-0.0^{}$&$-0.02^{}$&\cellcolor{blue!25}$0.1^{*}$&$0.0^{}$&$-0.02^{}$&$-0.02^{}$&$0.08^{*}$&$-0.02^{*}$&$0.03^{}$\\
GBR&$0.08^{*}$&$0.05^{}$&$0.0^{}$&$-0.02^{}$&$-0.02^{}$&$0.06^{}$&$0.02^{}$&$-0.02^{}$&$-0.01^{}$&$0.08^{*}$&$-0.02^{*}$&$0.03^{}$\\
GRC&$0.0^{}$&$-0.0^{}$&$-0.01^{}$&\cellcolor{blue!25}$0.07^{}$&$-0.01^{}$&$0.05^{}$&$-0.02^{}$&$0.02^{}$&$-0.01^{}$&$0.04^{}$&$0.03^{}$&$0.01^{}$\\
IRL&$0.06^{}$&$0.03^{}$&$-0.01^{}$&$-0.01^{}$&$-0.02^{}$&$0.05^{}$&$0.0^{}$&$-0.02^{}$&$-0.01^{}$&$0.07^{*}$&$-0.03^{**}$&$0.02^{}$\\
ITA&$0.01^{}$&$0.0^{}$&$-0.02^{}$&$0.03^{}$&$-0.01^{}$&$0.05^{}$&\cellcolor{blue!25}$0.06^{}$&\cellcolor{blue!25}$0.04^{}$&\cellcolor{blue!25}$-0.0^{}$&$0.04^{}$&\cellcolor{blue!25}$0.05^{}$&$0.02^{}$\\
JPN&$0.05^{}$&$0.06^{}$&$0.02^{}$&$0.01^{}$&$-0.01^{}$&$0.09^{*}$&$0.05^{}$&$0.01^{}$&$-0.02^{}$&$0.07^{}$&$-0.01^{}$&$0.03^{}$\\
NLD&$0.07^{}$&$0.04^{}$&$-0.01^{}$&$-0.02^{}$&$-0.01^{}$&$0.07^{*}$&$0.02^{}$&$-0.01^{}$&$-0.01^{}$&$0.08^{*}$&$-0.02^{}$&$0.03^{}$\\
NOR&\cellcolor{blue!25}$0.1^{*}$&$0.03^{}$&$0.01^{}$&$-0.02^{}$&$-0.02^{}$&$0.07^{*}$&$0.03^{}$&$-0.01^{}$&$-0.02^{}$&$0.06^{}$&$-0.02^{*}$&$0.03^{}$\\
NZL&$0.08^{}$&$0.01^{}$&$-0.01^{}$&$-0.01^{}$&$-0.01^{}$&$0.07^{*}$&$0.04^{}$&$-0.01^{}$&$-0.01^{}$&$0.07^{*}$&$-0.02^{}$&$0.02^{}$\\
PRT&$0.02^{}$&$0.01^{}$&$-0.02^{*}$&$0.01^{}$&$-0.01^{}$&$0.08^{*}$&$0.0^{}$&$-0.0^{}$&$-0.02^{**}$&$0.06^{}$&$-0.01^{}$&$0.01^{}$\\
SWE&$0.1^{*}$&$0.06^{}$&$0.01^{}$&$-0.02^{}$&$-0.01^{}$&$0.09^{*}$&$0.03^{}$&$-0.01^{}$&$-0.02^{}$&$0.08^{*}$&$-0.03^{*}$&\cellcolor{blue!25}$0.04^{}$\\
TUR&$0.01^{}$&$-0.02^{*}$&$-0.01^{}$&$-0.02^{}$&$-0.01^{}$&$0.01^{}$&$0.02^{}$&$-0.04^{}$&$-0.01^{}$&$0.05^{}$&$0.01^{}$&$0.01^{}$\\
USA&$0.08^{*}$&$0.06^{}$&$0.01^{}$&$-0.02^{}$&\cellcolor{blue!25}$-0.0^{}$&$0.07^{*}$&$0.04^{}$&$-0.01^{}$&$-0.01^{}$&$0.08^{*}$&$-0.02^{}$&$0.04^{}$\\
\hline
Average&$0.06$&$0.04$&$-0.0$&$-0.01$&$-0.01$&$0.07$&$0.02$&$-0.01$&$-0.01$&$0.07$&$-0.01$&$0.0$ \\
            \hline
        \end{tabular}
    \end{center}\label{tab:euclidean}
    \caption*{\footnotesize{Shaded areas are the development modes with maximum coherence for each country. Stars indicate non-ambiguity significance. *:$p<0.1$, **:$p<0.05$}}
\end{table}

\end{document}